\documentclass[USenglish, oneside, twocolumn]{article}

\usepackage[utf8]{inputenc}
\usepackage[big]{dgruyter_NEW}

\DOI{foobar}

\cclogo{\includegraphics{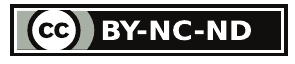}}

\usepackage{subcaption}
\usepackage{amsthm, amsmath, amssymb}
\usepackage{appendix}
\usepackage{hyperref}
\usepackage{booktabs}
\usepackage{multirow}
\usepackage{makecell}
\usepackage[table]{xcolor}

\usepackage{pifont}
\usepackage{bm}
\usepackage{color}
\usepackage{url}

\usepackage{graphicx}

\usepackage{pgfplots}
\pgfplotsset{compat=1.14}
\usepackage{filecontents}
\usepackage[all]{nowidow}
\usepackage{xspace}
\newcommand{\attack}{Var-CNN\xspace}

\begin{document}

\author*[1]{Sanjit Bhat}

\author[2]{David Lu}

\author[3]{Albert Kwon}

\author[4]{Srinivas Devadas}

\affil[1]{MIT PRIMES, E-mail: \href{mailto:sanjit.bhat@gmail.com}{sanjit.bhat@gmail.com}}

\affil[2]{MIT PRIMES, E-mail: \href{mailto:davidboxboro@gmail.com}{davidboxboro@gmail.com}}

\affil[3]{MIT, E-mail: \href{mailto:kwonal@mit.edu}{kwonal@mit.edu}}

\affil[4]{MIT, E-mail: \href{mailto:devadas@mit.edu}{devadas@mit.edu}}

\title{\huge Var-CNN: A Data-Efficient Website Fingerprinting Attack Based on
Deep Learning}

\runningtitle{Var-CNN}

\begin{abstract}
{In recent years, there have been several works that use website fingerprinting
techniques to enable a local adversary to determine which website a Tor user
visits. While the current state-of-the-art attack, which uses deep learning,
outperforms prior art with medium to large amounts of data, it attains marginal
to no accuracy improvements when both use small amounts of training data.
In this work, we propose \attack, a website fingerprinting attack that leverages deep learning techniques along with
novel insights specific to packet sequence classification.
In open-world settings with large amounts of data, \attack attains
over 1\% higher true positive rate
(TPR) than state-of-the-art attacks while achieving $4\times$ lower false
positive rate (FPR). Var-CNN's improvements are especially notable in
low-data scenarios, where it reduces the FPR of prior art by 3.12\%
while increasing the TPR by 13\%. Overall, insights used to develop Var-CNN
can be applied to future deep learning based attacks,
and substantially reduce the amount of training
data needed to perform a successful website fingerprinting attack. This
shortens the time needed for data collection and lowers the likelihood of
having data staleness issues.}
\end{abstract}


\journalname{Proceedings on Privacy Enhancing Technologies}
\DOI{10.2478/popets-2019-0070}
\startpage{292}
\received{2019-02-28}
\revised{2019-06-15}
\accepted{2019-06-16}

\journalyear{2019}
\journalvolume{}
\journalissue{4}

\maketitle

\section{Introduction}
\label{sec:intro}

Due to increases in mass surveillance and other attacks on privacy,
many Internet users have turned to Tor~\cite{tor} to protect their anonymity.
Over the years, Tor has grown to over 6,000 volunteer
servers and 4 million daily users~\cite{tor_metric}. Tor protects
its users' identities by routing each packet through a number of
Tor servers. Each server learns only the immediate hop before
and after itself, and as a result, no single server learns both the
identity of the user and the destination of the packet.

Unfortunately, Tor does not provide anonymity against
a powerful global adversary who can monitor a significant
portion of the traffic due to traffic analysis attacks.
In such attacks,
the adversary monitors traffic entering and leaving
the Tor network. Then, she uses traffic patterns such as packet sequences
to correlate packets across the two ends of the network
and determine the identities of the two communicating parties.
Recently, a variant of the traffic analysis attack called the
\emph{website fingerprinting} (WF) attack allows
an adversary who observes only the connection between the user and the
Tor network to identify which website the user visits.
To do so, the adversary first learns the traffic patterns of
certain websites, creating a unique digital fingerprint for each site.
Then, the adversary compares these fingerprints
to a user's network traffic (typically using machine learning algorithms)
to determine which website the user visits.

Most prior WF
attacks~\cite{cheng98,sun02,hintz03,liberatore06,herrmann09,bissias06,
  lu10,wang14,panchenko16,hayes16,wang13,panchenko11,cai12}
use manually extracted features to carry out the attack.
That is, the attacker carefully studies different protocols
such as HTTP and Tor, and manually determines features
that could potentially identify a website from a network trace
(e.g., the total number of packets and the total transmission time).
While these attacks achieved good accuracy in many settings,
it remains difficult to generalize the attacks across different protocols
or reason about their strengths, since the attacker
must manually specify the features.



Recently, deep learning neural networks have become the state-of-the-art
machine learning technique in several different domains such as
Computer Vision and Natural Language Processing~\cite{lecun15}.
While they can take in standard manually extracted features, one of their main
advantages is their ability to automatically
learn salient features just by analyzing input and output pairs.
This provides them the opportunity to discover more powerful features
than previously known, achieving higher accuracy. Current
state-of-the-art WF attacks~\cite{rimmer17,sirinam18} indeed
outperform prior manual feature extraction attacks in settings
with sufficient amounts of data.

One significant drawback of deep learning, however, is that it generally
requires a large amount of training data. As a result, deep learning WF
attacks have struggled to outperform traditional WF attacks in \emph{low-data
scenarios}. The current best deep learning WF
model by Sirinam et al., for example, achieves similar results as CUMUL,
the state-of-the-art manual feature extraction attack, when both use small
amounts of training data~\cite{sirinam18}.
Performance issues in low-data scenarios can be a serious issue for WF attacks:
since website traces change quickly (i.e., in the span of a few hours to a few
days), the attacker must frequently update her database of traces to match user
traffic~\cite{juarez14}.
Consequently, the adversary naturally needs to be stronger
to collect the larger amounts of traces required
for deep learning based attacks,
which weakens the attack in practice.


\subsection{Our Contributions}
In this work, we focus on answering the following question:

\textit{Can deep learning WF attacks achieve improved accuracy
over prior state-of-the-art WF attacks even with small amounts of training
data?}

We answer this question affirmatively using our new attack called \attack,
a semi-automated feature extraction (i.e., benefitting from both manual
feature extraction and automated feature extraction) WF attack based on deep
learning. To the best of our knowledge, \attack is the
first deep learning WF attack specifically tailored to network packet sequence
classification.
\attack's base architecture uses ResNets~\cite{resnet}, state-of-the-art
convolutional neural networks (CNNs) in Computer Vision.
Beyond the standard model architecture, 
\attack uses
the following key insights about the nature of packet sequences to achieve high performance:

\begin{enumerate}
  \item
    Packet sequences have a fundamentally different structure than images
    or other datasets traditionally classified using CNNs.
    For instance, numerical digits in the MNIST~\cite{lecun1998} dataset are
    composed of individual edges that can be detected by small feature detectors.
    Packets, on the other hand, typically have a more intertwined, global relationship
    (e.g., packets at the beginning of a trace can cause a large ripple
    effect throughout the whole trace).
    Unfortunately, simply increasing the size of individual feature detectors
    results in an unwieldy increase in the computational and memory overhead.
    Instead, we explore the use of
    \emph{dilated causal convolutions}, a variation found in Audio
    Synthesis~\cite{oord16} and Computational Biology~\cite{gupta17} research,
    to exponentially increase feature detector size without increasing runtime
    (\S\ref{sec:dilations}).

  \item
    Unlike Computer Vision model inputs, which are highly abstract, network
    packet sequences naturally leak some \emph{cumulative statistical information}
    such as the total number of packets, total time, etc.
    Although these manually extracted cumulative features achieve poor
    standalone accuracy, we overcome this problem by combining them with
    our dilated ResNet during training rather than after. This allows
    \attack to be the first deep learning WF attack that combines both manually
    extracted and automatically extracted features to improve
    the model's overall performance (\S\ref{sec:metadata}).

  \item
    To the best of our knowledge, packet timing information has only been used
    minimally in prior art WF attacks.
    In contrast, the input generality and power from ResNets and dilated
    causal convolutions allow us to
    retain performance under a domain change from direction information to
    \emph{packet timing information},
    showing that timing leaks a significant amount of information
    (\S\ref{sec:inter_time}).
    Moreover, we show
    that combining the direction and timing information of a packet
    sequence results in a more accurate overall model (\S\ref{sec:ensemble}).
\end{enumerate}

By incorporating the above three insights into \attack, we
achieve substantial improvements over prior art in both the open- and closed-world
in every experimental scenario tested. For instance, with 900 monitored sites and 2500 monitored
traces per site in our largest closed-world, \attack improves prior art
accuracy from 96.5\% to 98.8\%. In our largest open-world setting tested,
\attack attains over a 1\% better true positive rate (TPR) than prior art
while achieving a $4\times$ lower false positive rate (FPR).

In addition, Var-CNN has significant improvements over prior art in settings
with small amounts of training data, making the WF attack easier to carry out by weaker
attackers. For instance, with 100 traces for each of 100 monitored sites in a
low-data closed-world, \attack achieves 97.8\% accuracy, whereas it would
take prior art $5\times$ as much training data to achieve a comparable accuracy
of 98.1\%. In a low-data open-world with just 60 monitored traces and 6000
unmonitored training traces, \attack reduces prior-art FPR by 3.12\% while
increasing TPR by 13\%.
In \S\ref{sec:why}, we provide an intuitive explanation for why
\attack works well in these settings. Moreover, although we used ResNets here
as our baseline CNN architecture, many of our insights are
architecture-independent and thus can be applied to any future deep
learning-based attack.

\section{Background and threat model}
\label{sec:background}

\begin{figure}[!tb]
  \centering
  \includegraphics[width=0.45\textwidth]{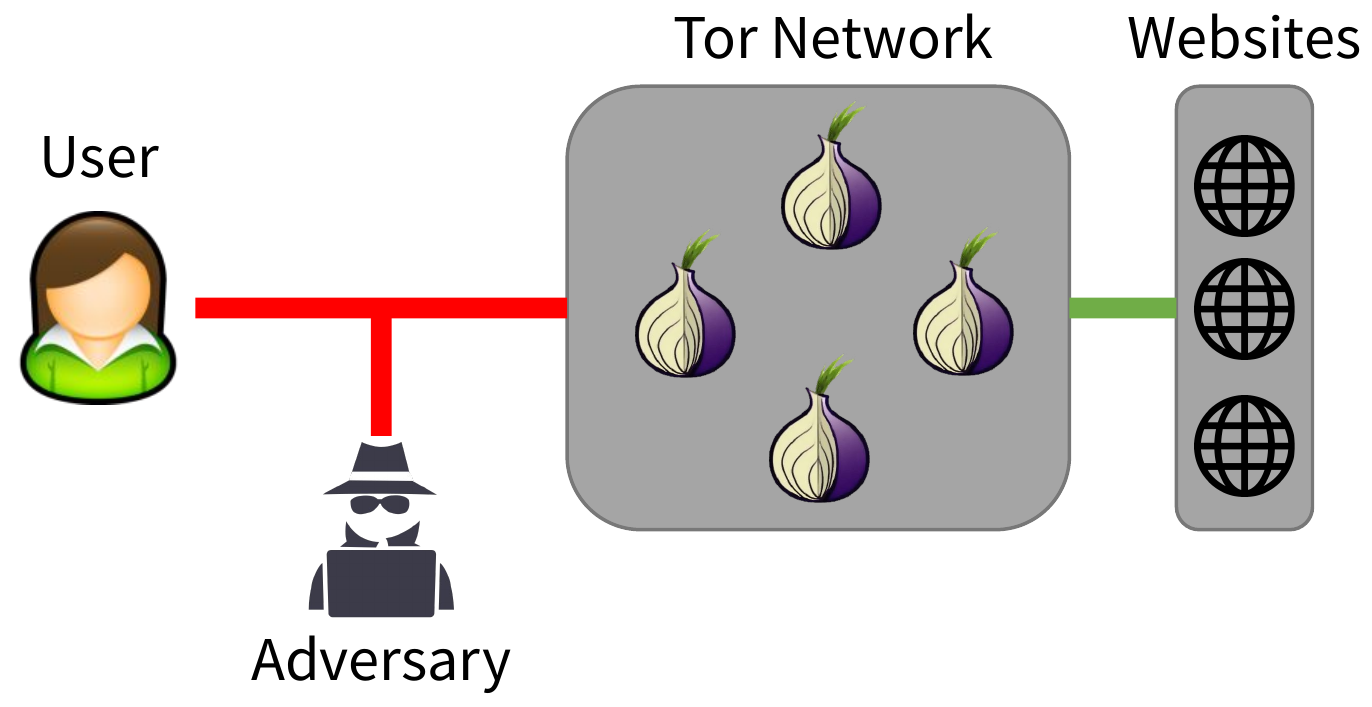}
  \caption{The website fingerprinting threat model.}
  \label{fig:threat_model}
\end{figure}

Tor~\cite{tor} consists of decentralized volunteer servers that relay
their users' packets without any delays or cover traffic.
While this has allowed Tor to scale to a large number of users,
this also enables adversaries monitoring traffic entering
and leaving the Tor network to potentially deanonymize users.
In particular, the website fingerprinting (WF) threat model allows an adversary
to monitor just the connection between a user and the network, as shown in Figure \ref{fig:threat_model}.
The adversary is a passive observer, meaning she will not drop, modify,
or insert any packets. Examples of such adversaries include
routers, internet service providers, autonomous services, and compromised Tor servers.

The adversary is interested in identifying visitors of a number of websites,
which we call \emph{monitored} websites; we call all other websites
\emph{unmonitored} websites.
The adversary visits the websites on her own and
creates a database of \emph{traces}, sequences of packets and
their timestamps generated while visiting a website. From here on, we refer
to whether a packet was incoming or outgoing as direction data and the
time delay between two consecutive packets as time data. Once the database is created, the
adversary monitors and collects users' traces and uses the database to classify
them as belonging to either the monitored or unmonitored set of websites.

\subsection{Attacker settings}
We consider two different attacker settings, closed-world and open-world.

\textbf{Closed-world.}
In this setting, we assume that users only visit
a well-known set of \emph{monitored websites}.
Here, the adversary trains on a number of traces from these sites
and aims to classify different traces from the training set
into one of the monitored websites.
We use \emph{accuracy} to define the effectiveness of
the attacker, which is simply the proportion of monitored traces correctly
identified. Although the closed-world is less realistic---it assumes the
adversary knows every site a user visits---it is a useful measure of a
classifier's ability to distinguish between websites.

\textbf{Open-world.}
In the real world, users can visit websites that the adversary does not know.
Open-world settings emulate this scenario by allowing users to visit
both monitored and unmonitored websites (i.e., websites that the adversary
does not deem sensitive). Just like in the closed-world, the
adversary trains on different traces from those being tested on. In
addition, the adversary can bias her classifier by training on
some number of unmonitored sites.
Though there is no overlap between unmonitored training and testing
sites, learning how to distinguish one set of unmonitored sites
often helps with others,
as we will see empirically in \S\ref{sec:ow_eval}.

In the open-world setting, we use three metrics to measure classification
performance, \textit{two-class true positive rate} (Two-TPR),
\textit{multi-class true positive rate} (Multi-TPR), and \textit{false positive
rate} (FPR). Two-TPR is the proportion of monitored traces correctly classified
as any monitored site, and it applies to an adversary who only cares about
identifying users who visit the monitored class in general. Multi-TPR is the
ratio of monitored traces correctly classified as a specific monitored site. An
adversary using this metric might assign different penalties to different
monitored sites. Finally, FPR, the ratio of unmonitored traces incorrectly
classified as a monitored site, measures the adversary's level of false
identification.

\subsection{Assumptions}
\label{sec:assumptions}
Prior WF attacks and \attack assume the following~\cite{juarez14}:

\begin{itemize}
\item
\textbf{Replicability.}
Given the uncertainty of real user conditions, the WF
attacker assumes her training data will be representative of actual Tor
traffic sequences.
\item
\textbf{Applicability.}
The adversary assumes she can apply the WF attack effectively in practice.
\end{itemize}

There are, however, several criticisms regarding these assumptions.
First, the replicability assumption may be too strong for a few reasons:
\begin{enumerate}
\item
Unless sites with dynamic content changes such as AJAX or Javascript are
adequately represented in the training set, the adversary would believe a
non-static page is actually static.
\item
Training on one Tor Browser version while a user uses a different version
could possibly result in different underlying protocols being used.
\item
Varying latency in Tor connections causes different inter-packet timings.
\item
The types of websites the adversary trains on may not match
the types of websites real users visit. For instance, the adversary
might train her classifier on Alexa's most popular
sites~\cite{alexa_topsites} whereas real Tor users might visit more private sites.
\end{enumerate}

Second, the applicability assumption faces the following issues:
\begin{enumerate}
\item
The adversary might not know when traffic from one website starts or stops, or
whether a user visits multiple websites at the same time.
\item
  Noise traffic (e.g., from listening to music or downloading a file in the background)
  might confuse an attacker.
\end{enumerate}

While some work has been done in applicability such as learning how
to split traces and remove background noise~\cite{wang16}, there
remain important problems in replicability due to the
privacy concerns of collecting realistic datasets from actual Tor users.
Our work does touch upon the replicability concern of data freshness with
dynamic websites, showing that deep learning attacks can perform well
with small amounts of training data. However, we acknowledge that,
similar to most prior work on WF attacks~\cite{panchenko11, dyer12, cai12, wang13, wang14,
panchenko16, hayes16}, there still exist replicability and applicability
assumptions
that could make the WF attack less powerful in practice.

\section{Related work}
\label{sec:related}
We now describe prior work in greater detail.

\subsection{Manual feature extraction attacks}
In the past, several WF attacks with manually extracted features have been
proposed, each directing attention towards the susceptible
components of the protocol studied~\cite{cheng98,sun02,hintz03,liberatore06,herrmann09,bissias06,lu10,panchenko11,dyer12,cai12,wang13,wang14,panchenko16,hayes16}.
For example, early work used weaknesses in HTTP 1.0 to take
advantage of distinct resource length leakage,
such as the size of images, scripts, and videos~\cite{cheng98, sun02, hintz03}.
Subsequent protocols hid resource lengths,
so later attacks focused instead on extracting information from packet
lengths leaked by HTTP 1.1, VPNs, and
SSH tunneling~\cite{liberatore06, herrmann09, bissias06, lu10}.
Since then, Tor and other anonymous networks that hide packet lengths have emerged.
Consequently, attacks in the last few years have focused on using a broad set of manually extracted
features~\cite{panchenko11, dyer12, cai12, wang13}.
For example, Wang et al.~used a modified $k$-Nearest Neighbors ($k$-NN)
classifier with a weight adjustment system to effectively train on a wide
feature set~\cite{wang14}, including packet lengths, packet orderings,
packet concentrations, and bursts.
Panchenko et al.~developed CUMUL, a Support Vector Machine (SVM) that mainly
relied on cumulative packet length features~\cite{panchenko16}.
Finally, Hayes et al.~in their $k$-FP attack used a slightly smaller feature
set than Wang et al.~and fed it into a Random Forest classifier, an
ensemble of Decision Trees~\cite{hayes16}. After training the Random Forest,
they fed the output into a vanilla $k$-NN classifier to control TPR and FPR.
Currently, Panchenko et al.'s CUMUL~\cite{panchenko16} is the best performing
manual feature extraction attack in vanilla WF settings, achieving
97.3\% accuracy in a small closed-world with a large number of
traces~\cite{sirinam18}.

While manual feature extraction attacks work well in some settings, they are
fundamentally restricted by their feature set. For
instance, Hayes et al.~pointed out that since CUMUL primarily
relies on packet ordering, it suffers significant decreases in
accuracy against simple defenses that perturb this information~\cite{hayes16}.
In addition, even though a broad feature set as in Hayes et al.'s
$k$-FP~\cite{hayes16} and Wang et al.'s $k$-NN~\cite{wang14} mitigates
information loss to a defense, these attacks are still only as good as their
feature set. For instance, while neither could effectively use timing, our
model achieves a non-trivial increase in performance when incorporating timing.
This exemplifies how difficult it can be for humans to manually
extract features.

\subsection{Automated feature extraction attacks}
Recently, a few authors have proposed work on automated feature
extraction (AFE) WF attacks using deep learning neural networks~\cite{abe16,
rimmer17}. Compared to traditional attacks, these attacks perform AFE over raw
input sequences, removing the need for feature
design. In an earlier work, Abe and Goto~\cite{abe16} used
a stacked-denoising autoencoder (SDAE), a neural network that tries to
create a compressed version of its input, to perform AFE.
Their model performed worse than Wang et al.'s $k$-NN
with manual features.
Later, Rimmer et al.~\cite{rimmer17} studied
preliminary applications of SDAEs, recurrent neural networks (RNNs), and
convolutional neural networks (CNNs)
and showed that AFE models can slightly outperform prior art when using
large amounts of training data.

The current state-of-the-art WF attack is Sirinam et al.'s~\cite{sirinam18}
Deep Fingerprinting (DF) attack, which uses a CNN architecture similar to
Simonyan et al.'s VGG model~\cite{simonyan14}.
DF showed improvements over prior art in both closed- and open-world settings.
However, with small amounts of training data (i.e.,
less than 200 traces in a closed-world with 95 monitored sites),
DF achieved marginal to no improvements over prior art, with both attacks
at around 90\% accuracy when using 50 traces~\cite{sirinam18}.
Even so, to the best of our knowledge DF is currently the strongest attack to
date in all domains.
Thus, we note several key differences
between \attack and DF:
\begin{itemize}
\item
  We employ several novel insights specific to packet
  sequence classification including dilated causal convolutions, cumulative
  statistical information, and timing data (\S\ref{sec:description}).
\item
  We show that \attack outperforms
  DF in every open- and closed-world setting tested, regardless of the
  amount of data used (\S\ref{sec:att_eval}).
\item
  Our model has noticeable improvements in settings with small amounts of
  training data. This is significant because of the following reasons:
  \begin{enumerate}
    \item Deep learning models typically do not
      work well with smaller training sets.

    \item Smaller training sets result in faster training times.

    \item Smaller training sets directly correspond to
      less work for the adversary as she
      needs fewer resources and less time to collect a database.
  \end{enumerate}
  The last point in particular strengthens WF attacks
  by allowing weaker adversaries to launch
  successful WF attacks and by reducing the chance of data staleness.
\end{itemize}

\section{\attack: Model variations on CNN}
\label{sec:description}
We first give a short background on convolutional neural networks
and then present details of \attack.

\subsection{Convolutional neural networks}
Convolutional neural networks are part of a class of machine learning
techniques called deep learning.
Unlike traditional machine learning, deep learning has the expressive power to
automatically extract features from raw input data by using several
hidden layers and non-linear activation functions such as ReLUs~\cite{lecun15}.
In our application, we use a convolutional neural network (CNN) due to
its hierarchical abstraction of features and reuse of features through local
connectivity. CNNs, which consist of three basic types of layers, can express complex
relationships between locally-defined features to construct more abstract
features~\cite{krizhevsky2012}.

Specifically, \textit{convolutional layers} make use of input translational
invariance (i.e., a feature appearing in multiple places) to have
locally-connected filters convolve over the entire input and create
feature maps. By default, each filter spans a small \textit{receptive field},
which will be further discussed in \S\ref{sec:dilations}. \textit{Pooling
layers} combine adjacent activations from feature maps and
downsample their input. Finally, \textit{fully-connected layers} at the end of
the network do away with local connectivity to have every neuron connected to
every neuron in the prior layer. In recent CNNs~\cite{resnet, inception},
fully-connected layers have only been used in the final \textit{softmax} layer
since they introduce a large number of additional parameters. The softmax
layer, discussed in \S\ref{sec:confidence}, represents a probability
distribution over the set of possible classes and gives the network's
confidence in a certain prediction.

\begin{figure}[t]
	\centering
    \includegraphics[width=0.25\textwidth]{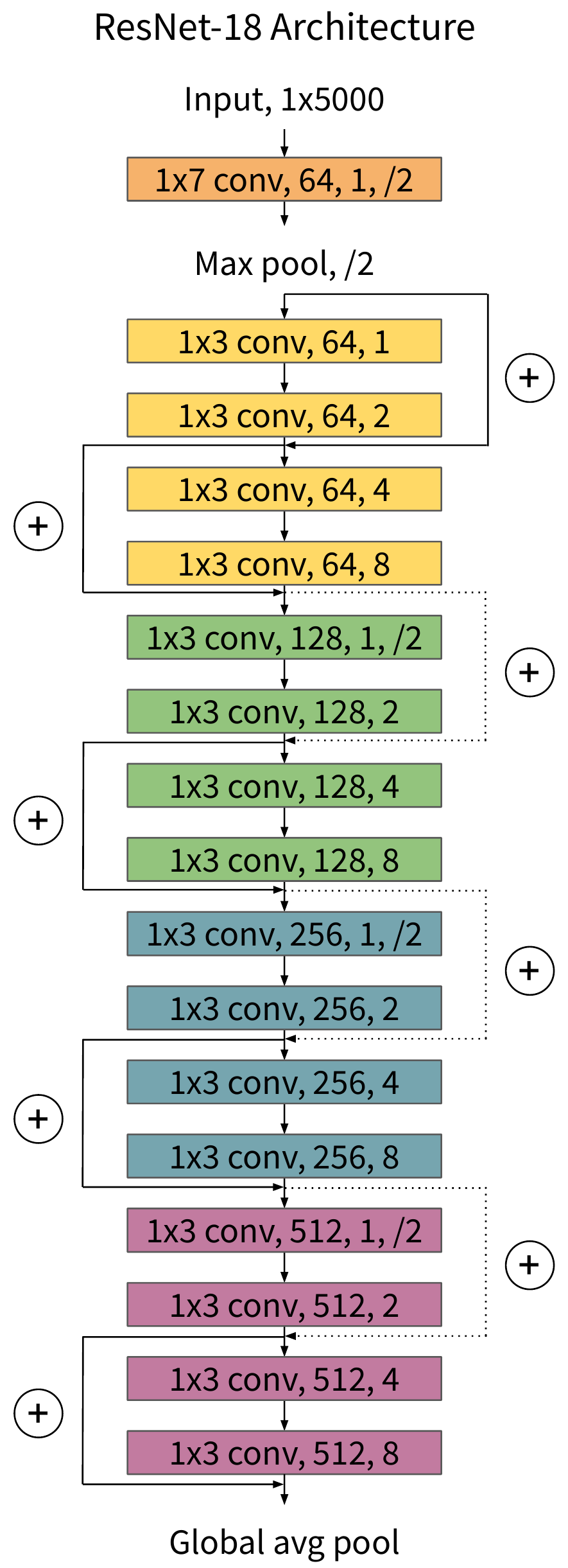}
    \caption{A graphical model of the \attack baseline CNN architecture, a
    ResNet-18, with dilated causal convolutions. The indices in each
    convolutional layer signify the kernel
    size, the number of filters, the dilation rate, and an optional /2 if that
    convolution downsamples its input by using a stride size of 2 instead of 1.
    Connections with a plus sign next to them indicate skip connections, which
    add the input of a block to its output. A dotted skip connection indicates
    that the input is downsampled before being added so that its dimensionality
    matches that of the block's output.}
	\label{fig:resnet}
\end{figure}

\subsection{\attack baseline architecture}
Our baseline CNN architecture (Figure~\ref{fig:resnet}) is based on
the state-of-the-art CNN for Computer Vision, ResNet~\cite{resnet}.
ResNet comes in several different sizes, and here, we use the smallest
variant with just 18 layers total to minimize training costs. ResNet-18 has 4
separate stages each consisting of 2 convolutional blocks. Each block contains
2 convolutional layers with batch normalization~\cite{ioffe15} and ReLU
non-linearity in-between. In addition, the key feature of ResNets that helps in
the optimization of larger networks is a residual ``skip'' connection between
the input and output of a block. The idea here is that deeper networks can be
hypothetically created from shallower networks by simply making deeper
blocks copy previous blocks.

For instance, consider a regular block that transforms its input $out = F(x)$,
where $F(x)$ is a series of convolution layers and nonlinearities. If the
aforementioned identity mappings are optimal,
then the block will have to fit the identity function,
something hard to do for a sequence of nonlinear layers. On the other hand, if
we explicitly  introduce the identity into each block, $out = F(x) + x$, it is
much easier for Stochastic Gradient Descent to fit the nonlinear block to
zero than to the identity function~\cite{resnet}. Therefore,
residual connections add the identity from each block to the output, easing
larger network optimization. This aids in higher-level feature
extraction and increases expressive power.

We train using the Adam optimizer, a variant of stochastic gradient descent,
to increase computational efficiency and accelerate convergence~\cite{kingma2014}.
By default, our ResNet-18 takes in packet direction information as input (i.e.,
1's and -1's that represent packets going to and away from the web server, respectively).
In the following sections, we present our augmentations to the baseline ResNet-18.

\subsection{Dilated causal convolutions}
\label{sec:dilations}
\begin{figure}[t]
    \includegraphics[width=0.5\textwidth]{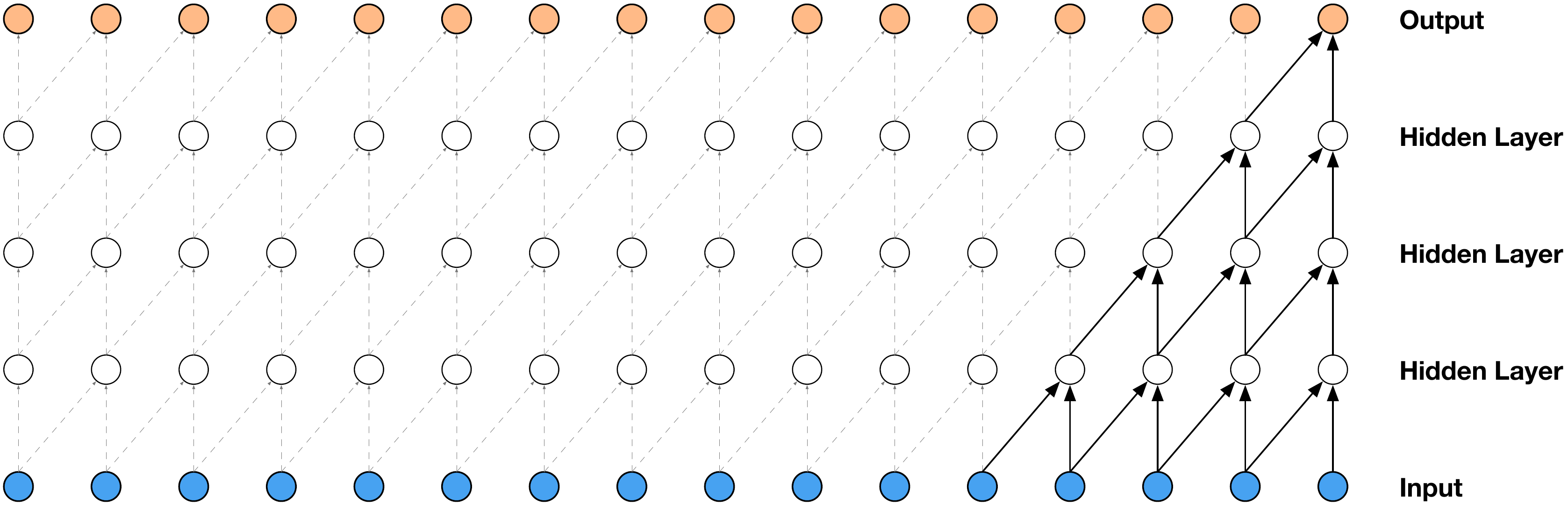}
    \caption{A pictorial representation of regular causal
    convolutions (figure replicated from Oord et al.~\cite{oord16}).
    Observe how the receptive field size increases
    linearly with the number of layers.}
	\label{fig:mc}
\end{figure}

\begin{figure}[t]
    \includegraphics[width=0.5\textwidth]
    {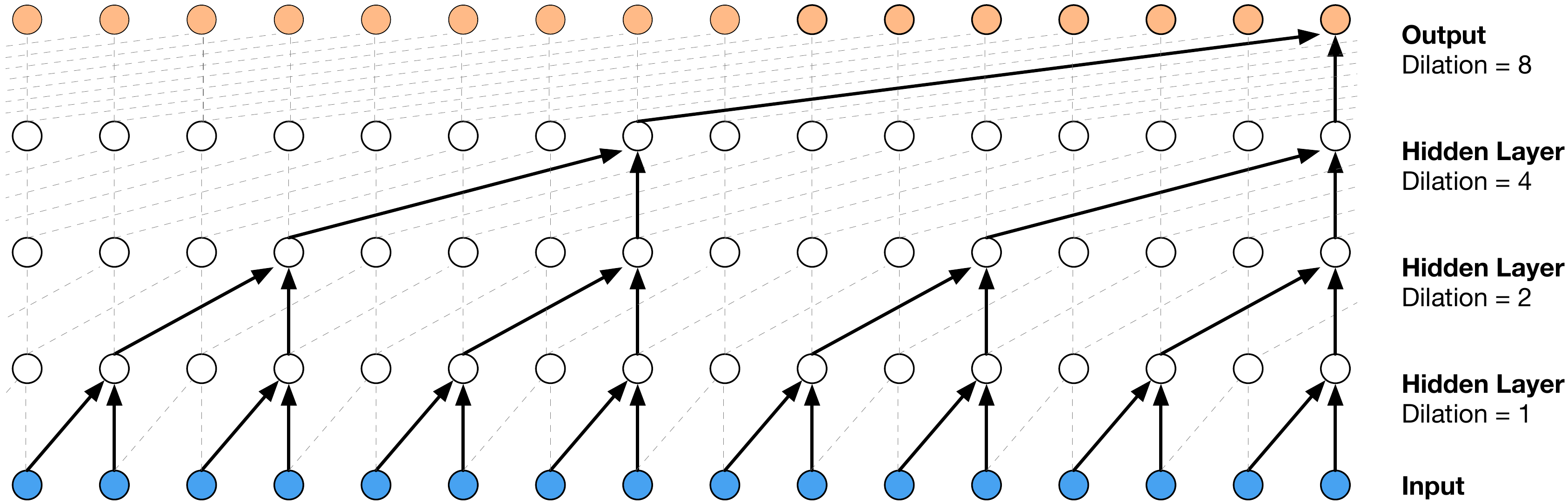}
    \caption{A pictorial representation of dilated causal
    convolutions (figure replicated from Oord et al.~\cite{oord16}).
    Each layer doubles the dilation rate, starting
    from a dilation rate of 1, or normal causal convolutions. This results in
    an exponential increase in receptive field size compared to a linear
    increase with normal causal convolutions (see Figure~\ref{fig:mc}).}
	\label{fig:dmc}
\end{figure}

With traditional convolutional operations, filters in each layer
convolve with their input volumes to produce feature maps.
Specifically, in ResNet-18, each filter has a small receptive field of
size 3, meaning that it can only operate on inputs that are at
most two apart from each other. Of course, higher-up layers operate on lower
layers, compounding receptive field, but this still results in each layer
having a receptive field that only grows linearly with the number of layers,
as shown in Figure~\ref{fig:mc}.

While a linear increase may be sufficient in image classification problems
where the inputs typically contain smaller and more locally-defined features,
packet sequences often have a much more long-term and intertwined structure.
For instance, consider classifying a human face.
Typically, the classifier would first detect smaller facial features
such as eyes, mouth, and ears, and combine these features
at a higher level to detect a face.
The smaller features happen within local spatial regions and can
thus be detected with linear receptive field increases. In contrast, when a
user contacts a server to download data, that might cause a cascading ripple
effect wherein several hundred streams of packets are sent before that one
transmission is completed. This insight about the long-term and
temporally-related nature of packet sequences motivates the following
exploration of dilated convolutions.

Strictly speaking, a larger increase in receptive field size could be
achieved through uniformly larger filters. However, in practice this results in
far more weight parameters and thus increases training time and memory.
RNNs such as Long Short-term Memory
Networks~\cite{hochreiter1997} have also traditionally provided support for
temporal data, but we found that the length of our packet sequence proved to be
prohibitively long for current RNN optimization techniques.\footnote{We
believe that the packet sequence is prohibitively long to train on RNNs
due to vanishing gradients, wherein the gradient updates
are essentially lost after backpropagating through such a long sequence.}

Instead, we build long-term temporal understanding into our CNN model by
utilizing \textit{dilated convolutions}~\cite{yu16, oord16}, convolutions
that skip inputs at a certain dilation rate.
Intuitively, instead of taking a fine-grain view of a small input region,
dilated convolutions allow the network to take a coarse,
wide view of the network, as shown in Figure~\ref{fig:dmc}. Since the
actual filter used in dilated convolutions is still the same as in
regular convolutions, the number of parameters and training cost do not
increase.

Our model doubles dilation rates in every convolutional layer until hitting
an upper-bound of dilation 8, upon which it cycles back to dilation 1 (i.e.,
dilation rates are \{1, 2, 4, 8, 1, 2, 4, 8, \ldots\}). Instead of a
linear increase in receptive field size, this results in an \emph{exponential}
increase, albeit for a small number of steps, as shown in Figure~\ref{fig:dmc}.
This technique has shown to be effective in Image Segmentation~\cite{yu16},
Audio Synthesis~\cite{oord16}, and Computational Biology~\cite{gupta17}.

Finally, as did Oord et al.~in their WaveNet model~\cite{oord16}, we also
combine dilated convolutions with causal convolutions, which simply restrict
every neuron to only look at neurons from previous timesteps. This helps
\attack better map the inherent temporal dependencies of packet sequences.
For instance, we found that
using dilated causal convolutions results in an increase in accuracy from 94\% to 96\%
in a closed-world of 100 monitored sites and 90 monitored instances.

\subsection{Cumulative statistical features}
\label{sec:metadata}
In addition to automatically extracting features from the raw data with
ResNet-18, we also provide seven basic cumulative features to the model.
These features include the total number of packets, the total number of
incoming packets, the total number of outgoing packets, the ratio of incoming
to total packets, the ratio of outgoing to total packets, the total
transmission time, and the average number of seconds to send each packet.

Based on our experiments,
a fully-connected layer with cumulative features as input achieves
only 35\% accuracy in a small closed-world (100 sites, 90 instances).
Due to the low accuracy, if we combined the metadata input and
ResNet-18 post-training by averaging their softmax outputs,
it would reduce the overall accuracy
(see Appendix~\ref{sec:model_design} for further details).
For example, in a small closed-world, adding metadata post-training results in an
accuracy \textit{decrease} in the dilated causal ResNet-18 from 96\% to 95\%.
Instead, we combine the fully-connected metadata layer
with the ResNet-18 during training by concatenating their outputs, as shown in
Figure~\ref{fig:ensemble}. Finally, after concatenation, the combined output is
sent through another fully-connected layer with Dropout
regularization~\cite{srivastava14} before going to the final softmax output.
Intuitively, this in-training ensemble scheme allows for the ResNet-18 AFE model to
automatically learn how to supplement its strong predictions with the weak
predictions from cumulative manually extracted features.
In practice, for a small closed-world, this results in an accuracy increase
from 96\% to 97\% for the dilated causal ResNet-18.
We note that while improvements seem marginal in this small closed-world setting,
they are amplified in more complex settings
where accuracies are lower (e.g., see \S\ref{sec:model_comp}).

\attack is the first WF attack to supplement automatically extracted
ResNet-18 features with cumulative manually extracted features.
Because of this, we consider \attack a semi-automated feature extraction model.

\subsection{Inter-packet timing}
\label{sec:inter_time}

To the best of our knowledge, low-level timing data (i.e., the timestep at
which a packet was sent or a close derivative of it) has never been
effectively used in prior art. For instance, Bissias et al.~\cite{bissias06}
used inter-packet timings in an early WF attack. However, compared to prior
art at the time using different features, their attack did not perform as
well~\cite{hayes16}. More recently, Wang et al.~\cite{wang14} used total
transmission time as one of their features, and Hayes et al.~\cite{hayes16} did
a feature importance study using inter-packet times. As shown by Hayes et al.,
however, these low-level time features ranked
40\textsuperscript{th}--70\textsuperscript{th} in feature
importance, making them essentially useless for classification.

Rather than dismiss packet timing, we applied the dilated ResNet-18 without
metadata (\S\ref{sec:dilations}) and observed some interesting
outcomes. Without changing any parameters and only
switching direction information with inter-packet times, the ResNet-18
with packet timing data achieved accuracy nearly comparable
to that of a ResNet-18 with direction data (96\% to 93\%, respectively,
in a closed-world with 100 sites and 90 instances).
Moreover, by combining it with the basic
cumulative features described in \S\ref{sec:metadata}, we were able to achieve
a 1\% accuracy improvement from 93\% to 94\% that is on par with what we
observed for the direction ResNet.

In contrast to prior work with manually extracted timing features, the high
accuracy of ResNet-18 with timing data indicates that packet timing
does leak a significant amount of information.
Apart from the suggested privacy leakage,
\attack with timing data highlights one of the key
benefits of AFE over manual feature extraction: performance under domain
shifts. Though no prior features have been discovered to effectively use
timing data, the ResNet-18 is both general enough to take in any sequence-like
input and powerful enough (with a strong ResNet architecture
and our augmentations) to perform well with these inputs.

As we will see in \S\ref{sec:model_comp}, ResNet-18 with time data is not
as accurate as ResNet-18 with direction data.
However, we describe a way to combine the two in \S\ref{sec:ensemble}
to improve the performance of the final classifier.

\subsection{Ensemble of timing and direction}
\label{sec:ensemble}
\begin{figure}[t]
	\centering
    \includegraphics[width=0.49\textwidth]{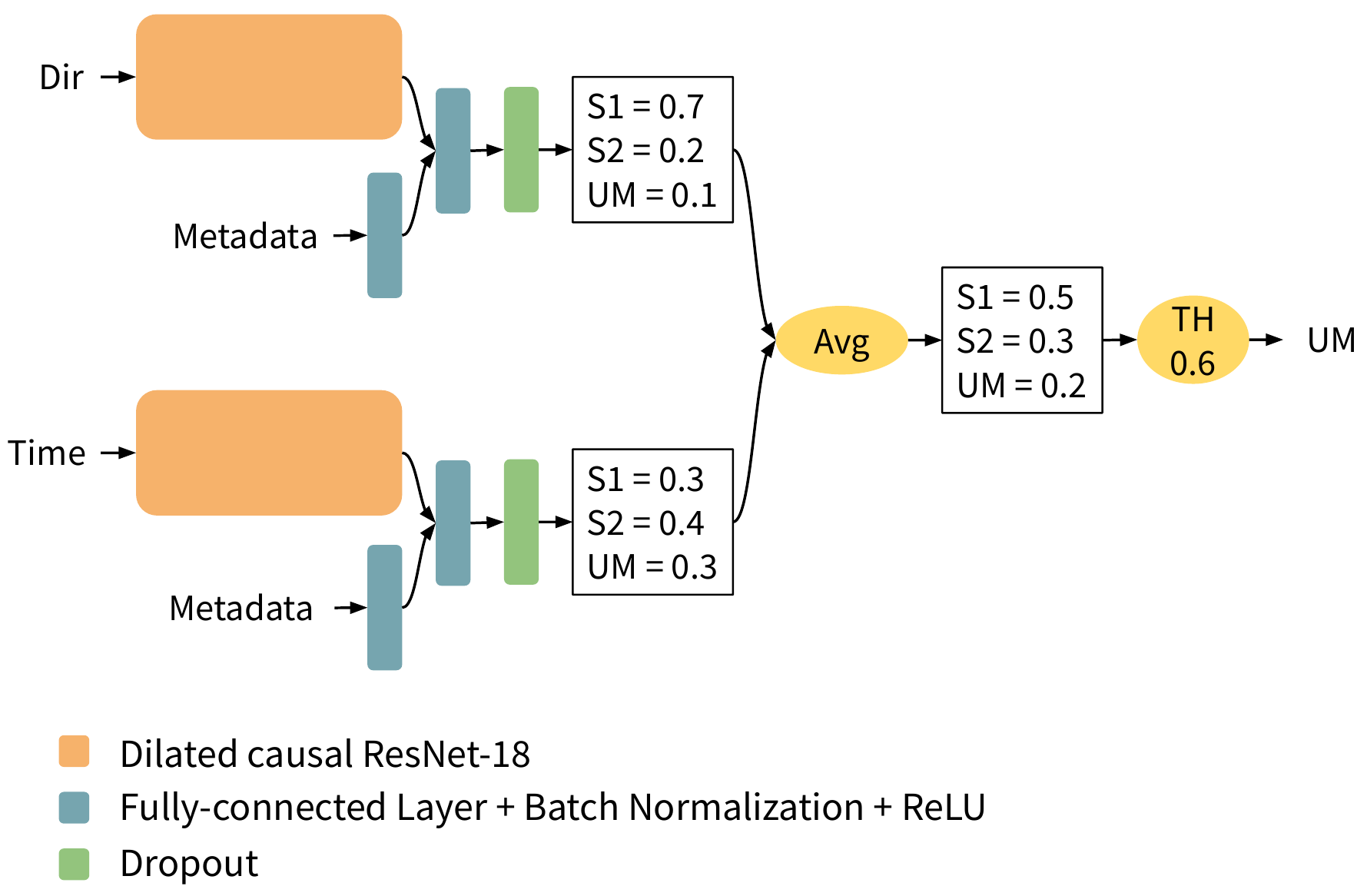}
    \caption{A graphical representation of our ensemble model with just three
    output classes, 2 monitored sites and the unmonitored (UM) class. Each
    of the direction and time models is trained separately in conjunction
    with metadata inputs. Finally, the two models' softmax layers are
    averaged post-training. TH signifies a minimum confidence threshold of a
    certain value over the averaged softmax outputs. The internals of the
    dilated causal ResNet-18 are shown in Figure~\ref{fig:resnet}.}
	\label{fig:ensemble}
\end{figure}

In \S\ref{sec:metadata}, we combined the ResNet-18 with
cumulative features in-training since the fully-connected layer with
cumulative features alone was far less accurate than the ResNet. In contrast,
both the time and direction ResNets with cumulative features are highly
accurate (as we will see in \S\ref{sec:model_comp}). Consequently, we found
that their accuracy actually dropped (from 97\% for the direction model to
95\% for the direction and time model in a small closed-world) when combining them during training.
This is most likely due to overfitting caused by training on essentially $2\times$
the number of parameters, which makes the underlying optimization problem much
more difficult (see Appendix~\ref{sec:model_design} for further details).

Instead, to effectively combine direction and timing, we take the arithmetic
mean of their softmax outputs after training each model separately.
This has the advantage of making each individual optimization procedure no
more parameter-intensive than the original, single-model optimizations. We
experimented with other averaging schemes (see Appendix~\ref{sec:model_design}
for further details) and found that even the optimal weighted average (i.e.,
the highest performing average over the test set)
always used each model equally, plus or minus a small $\epsilon$.
Since a simple arithmetic mean is near-optimal, the final \attack model
performs this over the outputs of the ResNet-18 direction and time models
with cumulative features,
as shown in Figure~\ref{fig:ensemble}.
In \S\ref{sec:model_comp}, we provide
empirical results that show how the averaging
scheme described here provides consistent improvements over the accuracy of
each individual model. 

\subsection{Confidence threshold}
\label{sec:confidence}
As the final step of our attack, we apply a post-training threshold on
the softmax probability output of the network (indicated by the ``TH'' block in
Figure~\ref{fig:ensemble}). If the output class probability is less than this
threshold, we change the predicted class to the unmonitored class.
Intuitively, this can be explained as defining
a certain minimum bound on model certainty before classifying a sample.
If a model is not certain about its classification as a monitored website,
we assume that the testing input was really an unmonitored website
that only partially matched a monitored site, and we classify it as unmonitored.

The threshold constraint also allows for direct control over TPR and FPR trade-off.
Prior manual feature extraction work achieved this using methods
such as classify-verify~\cite{juarez14}, using an additional classifier on top
of features outputted from a primary classifier~\cite{hayes16},
and changing the number of nearest neighbors in $k$-NN~\cite{wang14,hayes16}.
However, in the case of using a different model to perform final
classification, this results in increased computational time due to training
multiple models and decreased accuracy due to information loss
between the feature extractor and classifier. Additional schemes such as
classify-verify and adjusting the number of nearest neighbors require
re-training to adjust trade-offs.

In contrast, \attack allows us to easily adjust thresholds post-training.
With a threshold of 0, this is the equivalent of not applying any model constraint.
When the threshold is 1, we restrict the network to only classify websites
as monitored if it is 100\% certain, decreasing the number of true positives
and false positives. We discuss empirical results for the TPR-FPR trade-off in
\S\ref{sec:trade-off}. We further note that since confidence thresholds
result from having a final discrete softmax probability output,
both Rimmer et al.'s~\cite{rimmer17} and Sirinam et al.'s~\cite{sirinam18}
deep learning-based attacks have them.

\section{\attack evaluation}
\label{sec:att_eval}
In this section, we describe our experimental setup and evaluate \attack.

\subsection{Experimental setup}
\label{sec:procedures}
\subsubsection{Dataset}
Unfortunately, Sirinam et al.'s dataset was not available at the time of
writing. Instead, we evaluate our attack on the Rimmer
et al.~dataset~\cite{rimmer17}, which was publicly available. However, since
the public dataset does not have enough information to extract inter-packet
time and metadata, we processed their unfiltered dataset, which  we obtained upon
request.

This dataset has a total of 900 monitored sites each with 2,500 traces. For
open-world, there are a total of 500,000 additional unmonitored sites, each
with only 1 trace. Both the monitored and unmonitored pages were compiled from
the Alexa list of most popular sites~\cite{alexa_topsites}, and with over 2.75
million total traces, it is one of the largest and most up-to-date WF datasets
in existence.

For each of our three different models, we feed in a different set of features
representing a given trace.
The direction ResNet-18 takes in a set of 1's and -1's that represents the
direction of each packet.
The numbers 1 and -1 respectively denote outgoing packets
(which travel toward the web server) and incoming packets
(which travel toward the client).
The time ResNet-18 takes in a sequence of floats that represent the time
delay between when a current packet and the previous packet is sent.
The metadata model takes in seven floats (for the seven cumulative statistical features
described in \S\ref{sec:metadata}).
Since CNNs only take in a fixed-length input,
we pad and truncate each set of inputs to the direction and time models to the
first 5,000 values.
This is consistent with prior work~\cite{rimmer17, sirinam18} and strikes a
balance between the computational overhead of larger sequences and the
information loss of smaller sequences.

\subsubsection{Training/Validation/Test Split}
In all our experiments, we use the exact same training, validation,
and test data for both \attack and DF~\cite{sirinam18}.
Although manually extracted attacks used cross-validation on smaller
datasets~\cite{wang14, panchenko16, hayes16}, this is computationally
infeasible on our larger dataset. Instead, we follow the best practices of the
deep learning community and the practices of Sirinam et al.~\cite{sirinam18}
and split data into a training, validation, and test set. No random data split
favors any model because we use the same split for both models.

In all our settings, we used a random 10\% of all monitored traces (i.e., 10\% of the
traces from each website) for
testing and a random 5\% of the remaining 90\% of all monitored traces (regardless of
website) for validation. While the
number of unmonitored training and testing sites differs with every setting,
we use a random 5\% of all unmonitored training sites for validation.
There is no overlap between data used for training, validation, and testing.
In addition, note that we use the validation set to time learning rate decays
and stop model training (see Appendix~\ref{sec:val_params} for further details).

\subsubsection{Model Implementation}
To implement \attack, we use Keras~\cite{chollet2015} with the
TensorFlow~\cite{tensorflow15} backend. As Sirinam et al.'s code was not
available at the time of writing, we reimplemented their model with their guidance in
the same Keras and Tensorflow environment they used~\cite{sirinam18}. Given
that neural networks train several times faster on GPUs due to data
parallelism, we run our experiments on NVIDIA GTX 1080Ti workstations at our
organization with 11~GB of GPU memory.

\subsubsection{Hyperparameter Tuning}
To determine hyperparameters for \attack, we systematically go through each hyperparameter,
sweeping a certain range of values while keeping all other hyperparameters constant.
After finding a well-performing value, we fix it and then test other hyperparameters.
As mentioned in \S\ref{sec:description}, we measure performance using
accuracy on the test set for a small closed-world with 100 sites and 90 instances.
After selecting hyperparameters (e.g., architecture, regularization, and
starting learning rate) on this small closed-world, we fix them for all
subsequent open- and closed-world evaluations in \S\ref{sec:experiments}.
Our main reason for selecting hyperparameters once using a small dataset was to
increase the speed of our research, and we reported closed-world accuracies
in \S\ref{sec:description} simply to provide a metric for the performance
differences of various Var-CNN architectural choices, not as a point of
comparison against other attacks.

For further details on the specific parameters we used and for
additional information on model design choices that worked and did not work, see
Appendix~\ref{sec:model_design}.
Note that our parameter search was by no means extensive, especially for the
ResNet-18 with packet timing data. It is entirely possible that different \attack
configurations perform differently on direction data versus time data,
but we made a simplifying assumption to use the same configuration.
In addition, this showcases Var-CNN's input generality and high performance
under domain changes.

\subsection{Experimental results}
\label{sec:experiments}

In this section, we discuss our various experiments and their implications.

\subsubsection{Comparison of \attack variants}
\label{sec:model_comp}
\begin{figure}[t]
	\centering
    \includegraphics[scale=0.46]{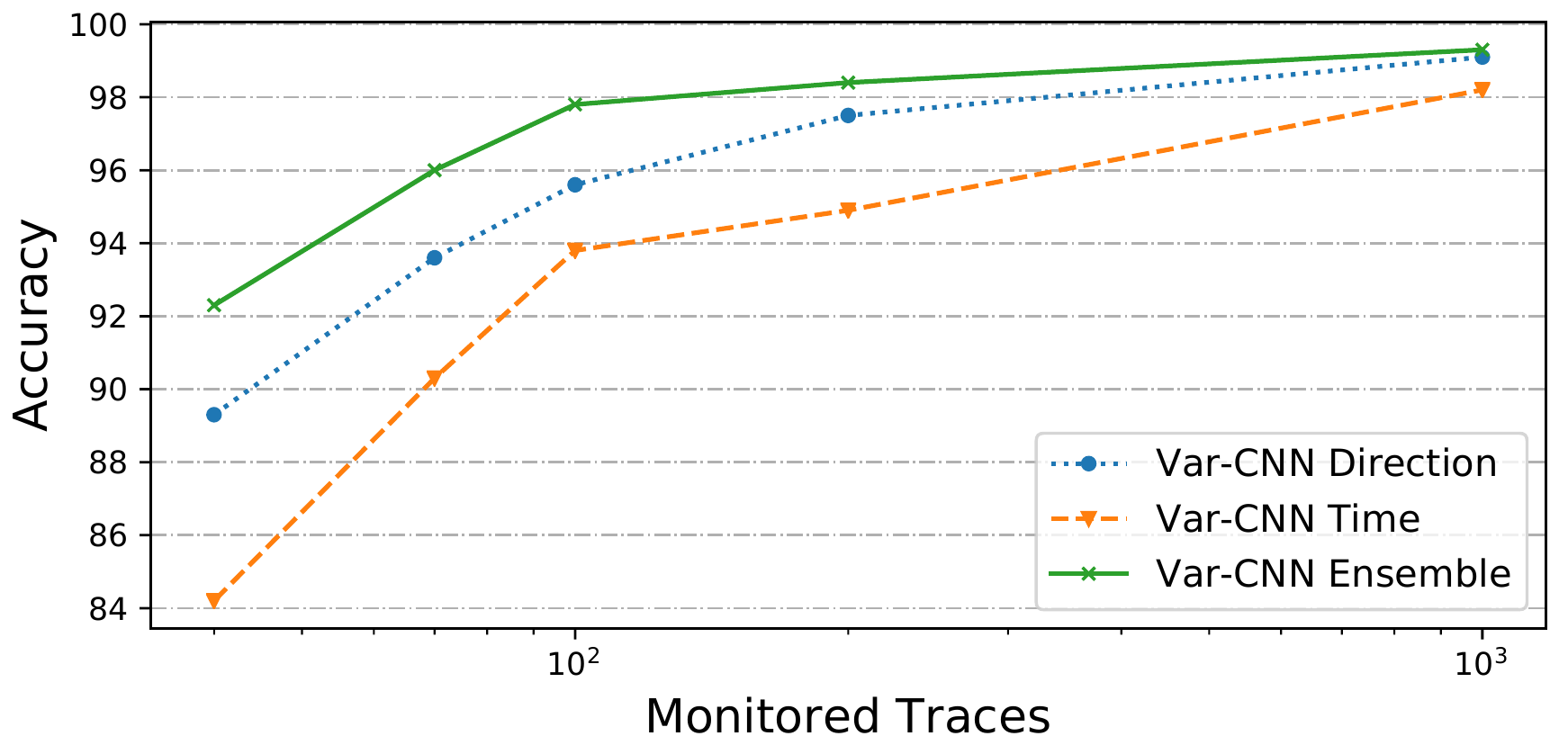}
    \caption{Closed-world accuracies of \attack direction, \attack time,
    and \attack ensemble. Experiments are run with 100 monitored sites and a
    varying number of traces per site. Accuracies are in \% and the x-axis is
    logarithmically scaled.}
	\label{fig:model_comp}
\end{figure}

We first compare the three different configurations of Var-CNN: direction,
time, and ensemble.

Figure~\ref{fig:model_comp} shows accuracies for these three
configurations as we change the number of monitored traces in a closed-world
setting. First, we observe that while the time model
always gets less accuracy than the direction model, both are still highly
comparable. For instance, with 1000 traces, the accuracy difference is only
0.9\%, meaning both models are highly accurate. While prior art was
unable to effectively use low-level timing data (\S\ref{sec:inter_time}),
we show that AFE can take advantage of this information.
\attack effectively classifies websites solely based on timing data
without any major modifications.

Second, the ensemble model combining direction and time always has higher
accuracy than either of its constituents. While this may seem obvious at
first---two models must always be better than one---it suggests that 
the two models
seem to complement each other's predictions. If both were highly confident on
the exact same predictions, then the ensemble created by averaging their
outputs would not improve. Instead, it appears that there are some
situations where one model is more confident than the other, and vice-versa.
These situations average out such that the ensemble model produces better
overall predictions than either of its constituents.

\subsubsection{TPR-FPR trade-off}
\label{sec:trade-off}
\begin{figure}[t]
	\centering
    \includegraphics[scale=0.46]{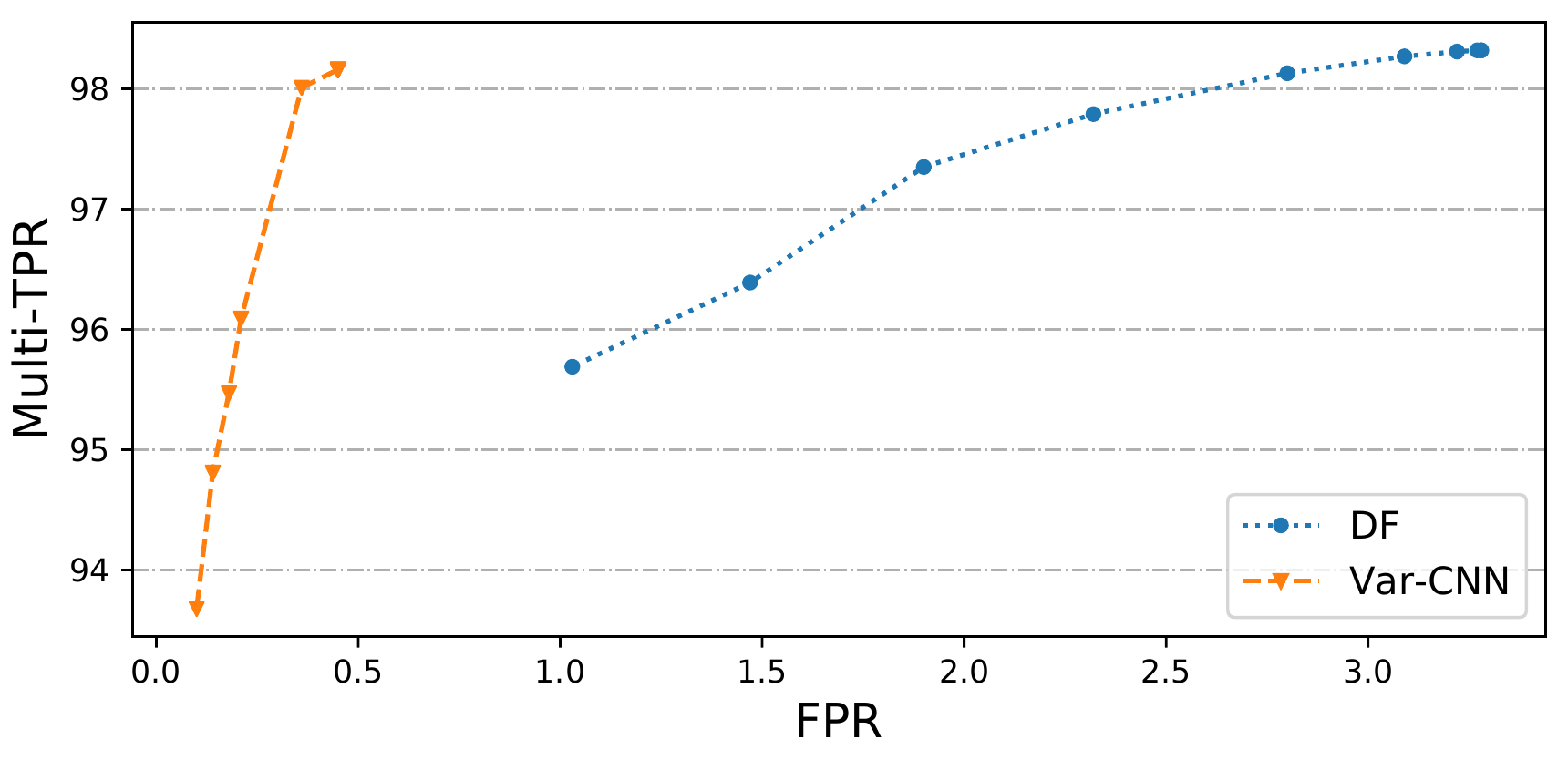}
    \caption{The TPR and FPR trade-off for an open-world setting with 100
    monitored sites, 1,500 monitored traces, 150,000 unmonitored training sites,
    and 100,000 unmonitored testing sites. Multi-TPR and FPR are in \% and the
    upper left corner is the best. Data points farther up and to the right use
    lower confidence thresholds.}
	\label{fig:trade-off}
\end{figure}

As discussed in \S\ref{sec:confidence}, \attack allows TPR-FPR trade-off
by changing the confidence threshold post-training. Using a higher
confidence threshold, the attacker can lower FPR by assuming that traces
predicted as a monitored site with low confidence are actually false positives.
This greedy scheme naturally causes true negatives, i.e., monitored
traces classified as unmonitored due to their relative dissimilarity to
other monitored traces. A similar technique can also be applied to DF,
and in Figure~\ref{fig:trade-off}, we plot this trade-off
for both attacks in our largest open-world setting tested.

As can be seen, both TPR and FPR decrease as the confidence threshold
increases and the graph goes to the bottom left. For instance, with a threshold
of 0.5, Var-CNN has a 98.01\% Multi-TPR and a 0.36\% FPR. As the threshold
increases to 0.9, Var-CNN's TPR goes down to 93.68\% but its FPR also goes down
to 0.10\%. With 100,000 unmonitored testing sites, this is a reduction of 260
false positives to only 100 false positives compared to the initial 360 false positives.

Another interesting trend is the difference between Var-CNN and DF
FPR for the same Multi-TPR. When both attacks use a threshold of 0,
for nearly the same TPR, DF has an FPR of 3.28\%, which is over $7\times$
that of Var-CNN at 0.45\%.
In addition, this difference gets larger for increasing Multi-TPR,
indicating that Var-CNN has significant FPR benefits when
the attacker needs to maximize TPR\@.

\subsubsection{Closed-world performance}
\label{sec:closed_world}
\begin{figure}[t]
	\centering
    \includegraphics[scale=0.46]{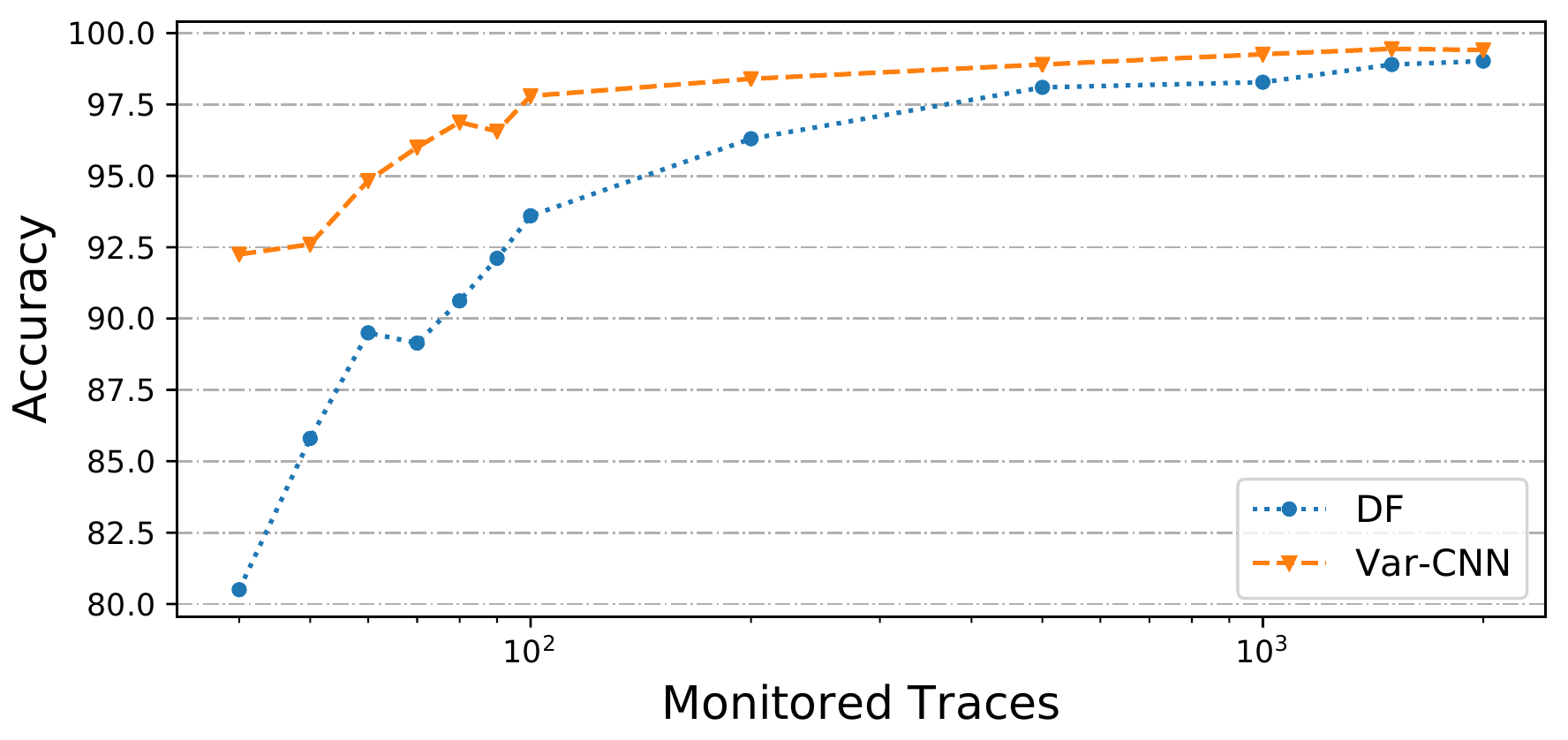}
    \caption{\attack and DF closed-world accuracy as a function of the
    number of monitored traces for each of 100 monitored sites. Results are
    in \% and the x-axis is logarithmically scaled.}
	\label{fig:closed_scaling}
\end{figure}

\begin{figure}[t]
	\centering
    \includegraphics[scale=0.46]{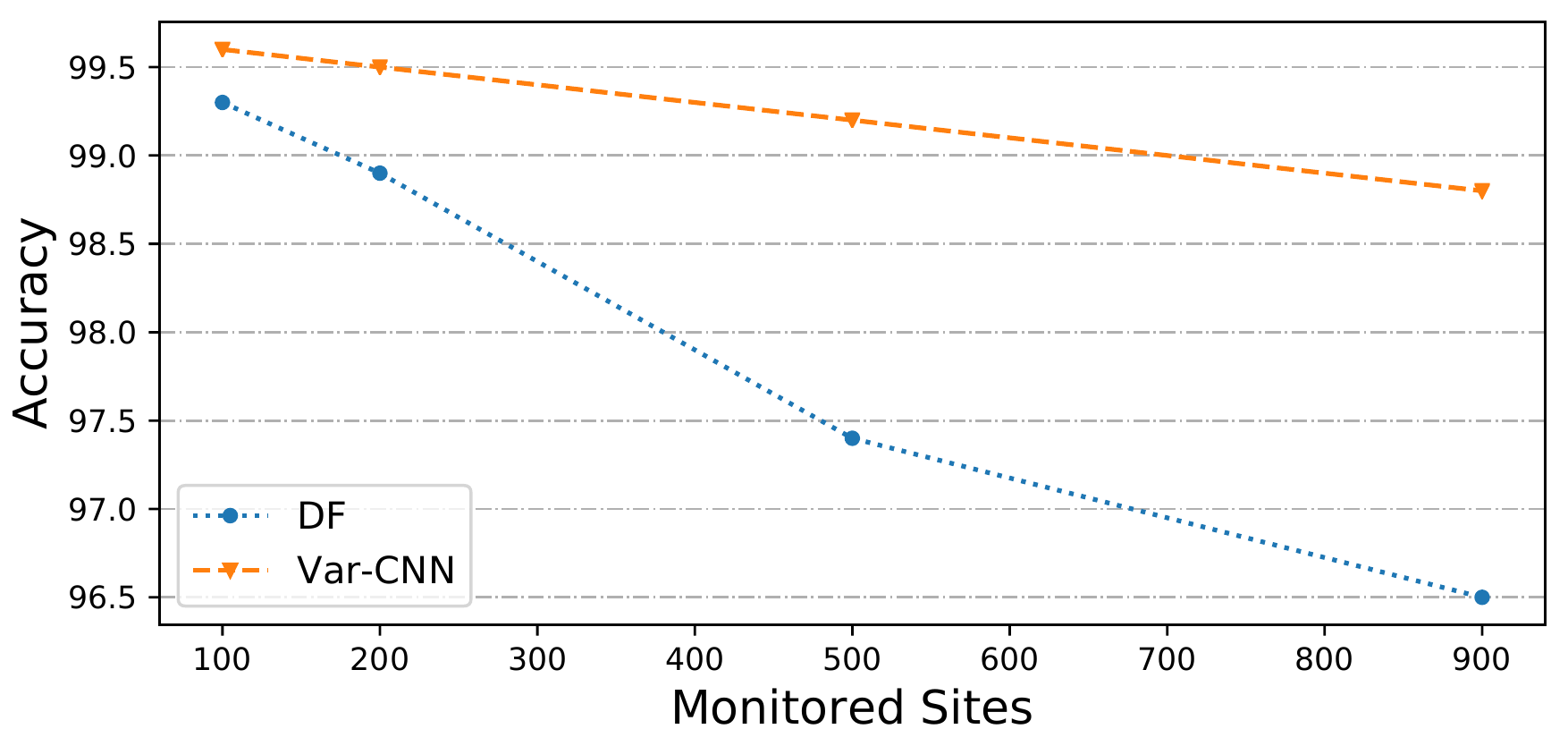}
    \caption{\attack and DF closed-world accuracies for larger settings with
    2500 monitored traces and a varying number of monitored sites. Results are
    in \%.}
	\label{fig:large_closed}
\end{figure}

We now consider closed-world experiments with varying amounts of
training data. Figure~\ref{fig:closed_scaling} shows \attack and Deep
Fingerprinting accuracy as the number of traces increases for each of 100
monitored sites. First, observe Var-CNN's high accuracy even in settings
with relatively small numbers of traces. For instance, with 100 traces, \attack
achieves 97.8\% accuracy while DF achieves 93.6\% accuracy. It
would take $5\times$ as much training data (i.e., 500 traces) for DF
to achieve a comparable accuracy of 98.1\%.

In general, this accuracy difference does not stay stagnant over time.
Rather, Var-CNN's accuracy improvements tend to increase as the number of
traces decreases.
For example, at 2000 traces, our largest amount tested, there
is a 0.39\% accuracy gap. This increases to 0.8\% with 500 traces, 2.1\% with
200 traces, 4.2\% with 100 traces, 6.8\% with 50 traces, and 11.75\% with 40 traces.

As shown in Figure~\ref{fig:large_closed}, \attack also has large accuracy
increases in closed-world settings with many sites and a large number of traces
per site. For instance, with just 100 sites and 2500 traces, both attacks
achieve near comparable accuracy with a difference of 0.3\%. However, as the
number of monitored sites increases with no additional traces, the gap between
\attack and DF accuracy
also increases to 0.6\% with 200 sites (99.5\% compared to 98.9\%),
1.8\% with 500 sites (99.2\% compared to 97.4\%), and 2.3\% with 900 sites
(98.8\% compared to 96.5\%). \S\ref{sec:why}
provides a unifying explanation for the increasing accuracy gap phenomenon in
both low-data and high-data scenarios.

\subsubsection{Open-world performance}
\label{sec:ow_eval}
\begin{figure}[t]
	\centering
	\begin{subfigure}[t]{0.46\textwidth}
        \includegraphics[width=1\linewidth]{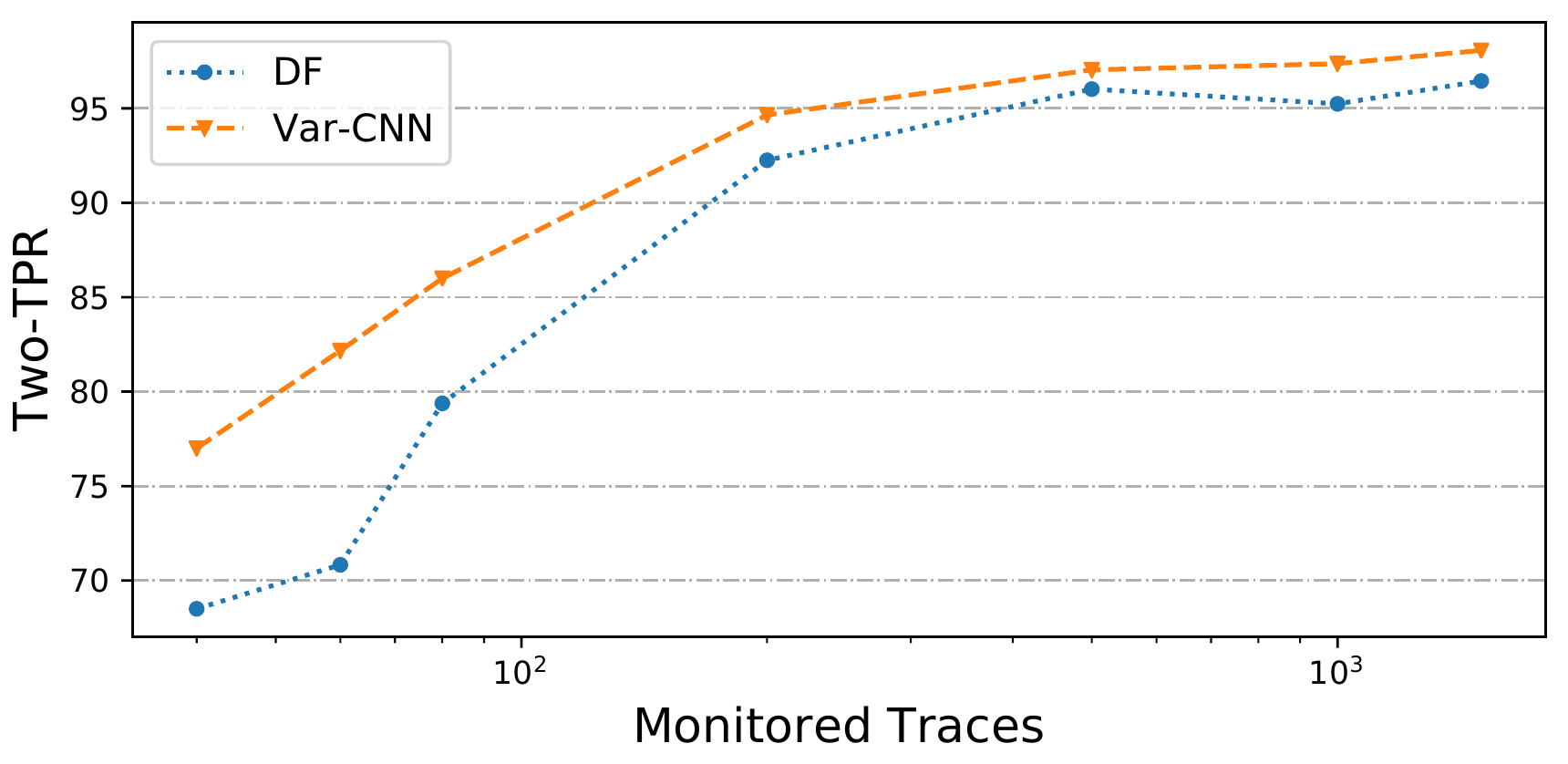}
        \captionsetup{justification=centering}
        \caption{}
        \label{fig:ow_two_tpr}
    \end{subfigure}

    \begin{subfigure}[t]{0.46\textwidth}
        \includegraphics[width=1\linewidth]{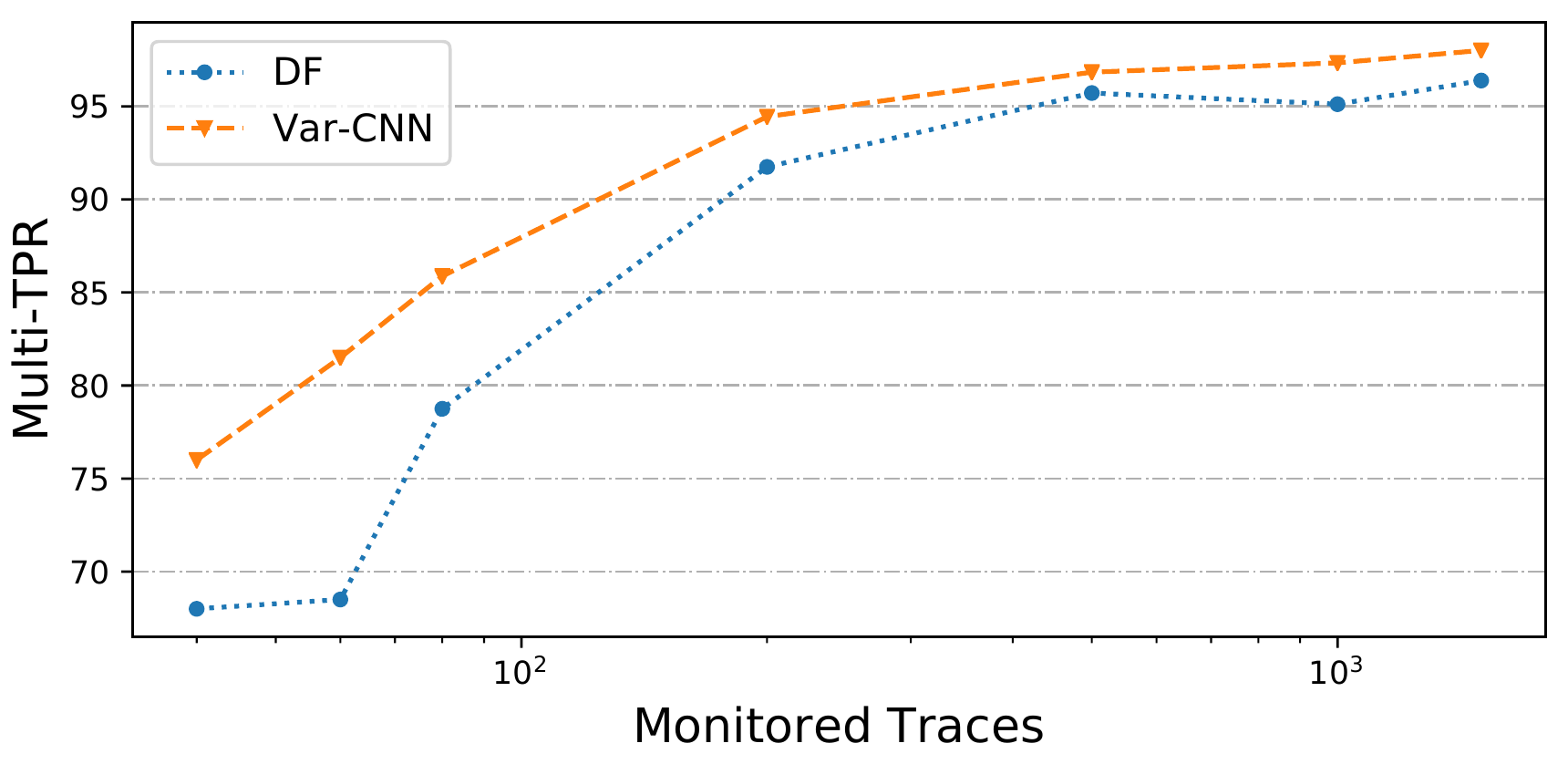}
        \captionsetup{justification=centering}
        \caption{}
        \label{fig:ow_multi_tpr}
    \end{subfigure}

    \begin{subfigure}[t]{0.46\textwidth}
        \includegraphics[width=1\linewidth]{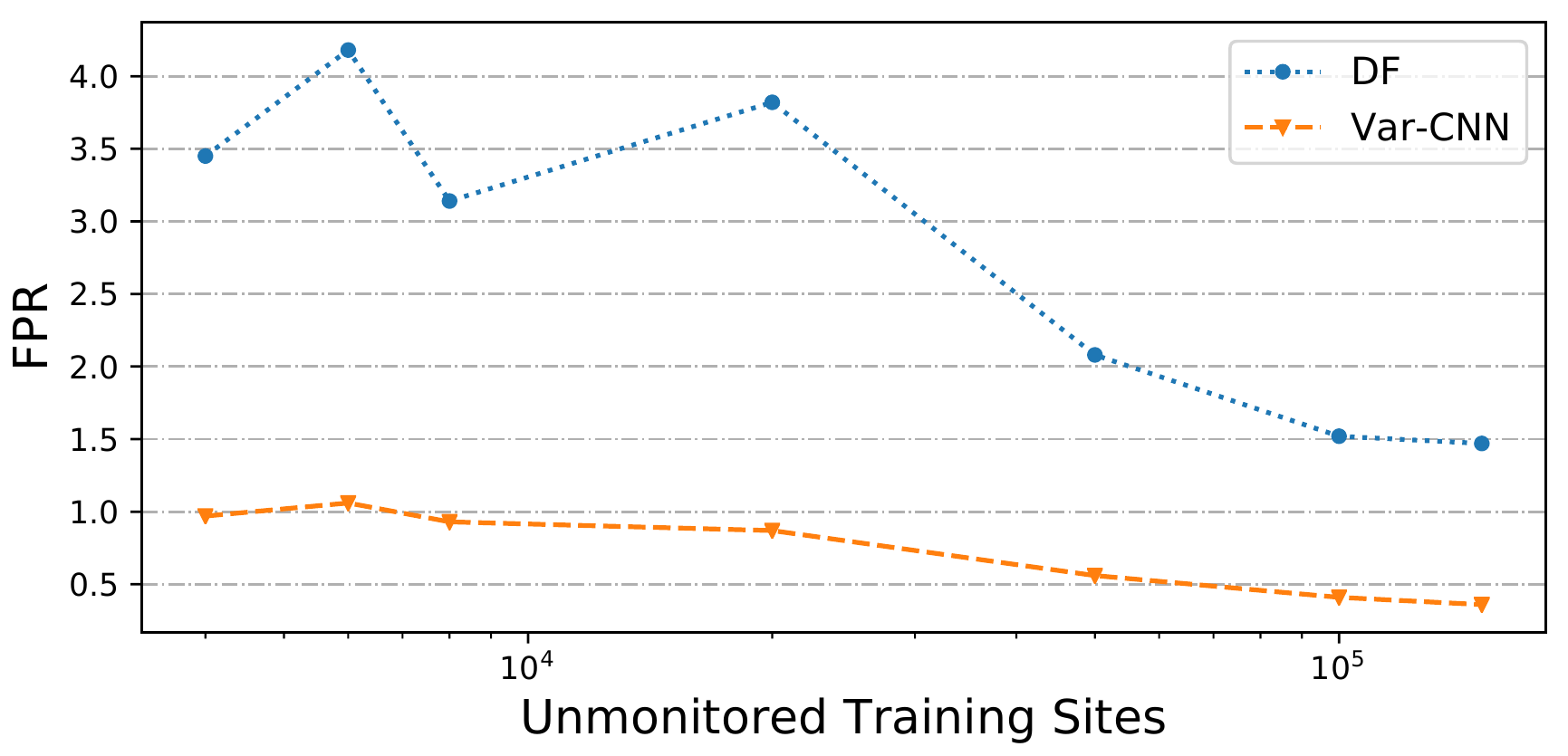}
        \captionsetup{justification=centering}
        \caption{}
        \label{fig:ow_fpr}
    \end{subfigure}

    \caption{\attack and DF's (a) Two-TPR, (b) Multi-TPR, and (c) FPR as a
    function of the attacker's data collection capabilities. Here, we use 100
    monitored sites and 100,000 unmonitored testing sites. For each setting, we
    use monitored traces and unmonitored training sites in a 1:100 ratio. We
    choose a confidence threshold of 0.5 for \attack and 0.8 for DF. Results
    are in \% and the x-axis is logarithmically scaled.}
\end{figure}

We now evaluate Var-CNN and DF in the
open-world setting. Here, both the number of monitored
traces per monitored site and the number of unmonitored training sites
affect TPR and FPR. Generally, more monitored traces leads to a
better knowledge of the monitored class and a higher TPR, while more
unmonitored training sites biases the attack model towards correctly separating
monitored from unmonitored, reducing FPR. Of course, these trends are
not exact, as strictly increasing the number of unmonitored training sites also
introduces noise in monitored classification, slightly reducing TPR.

We assume that the adversary can collect enough of both
types of traces to balance out possible negating effects.
Specifically, we quantify the attacker's data collection capabilities by
using a varying number of monitored traces and unmonitored training
sites in a 1:100 ratio. We now study how \attack and DF perform
under an adversary with varying data collection capabilities. For all
our experiments, we picked
confidence thresholds such that both attacks had a good balance of TPR and
FPR and used these across all data settings.

In large open-worlds where the attacker must separate
perhaps billions of unmonitored sites from monitored, false positives
are often the limiting factor~\cite{panchenko16}. For example, with a million
unmonitored traces, an FPR of even 0.1\% results in 1,000 false positives.
Observe in Figures~\ref{fig:ow_two_tpr},~\ref{fig:ow_multi_tpr},
and~\ref{fig:ow_fpr} that for both \attack and DF, Two-TPR
increases, Multi-TPR increases, and FPR decreases as the adversary's data
collection capabilities increase. Assuming the probability of an unmonitored
site falsely classified as a monitored site stays the same for an arbitrarily
sized open-world, an adversary can thus reduce false positives
and increase true positives by scaling the amount of training data.
Furthermore, \attack performs better than DF in
these high-data settings (i.e., settings with a large number of traces per trial),
with over $4\times$ lower FPR (1.47\% to 0.36\%),
over 1\% better Two-TPR (96.45\% to 98.07\%), and over 1\% better Multi-TPR
(96.39\% to 98.01\%).

While \attack outperforms DF in high-data settings, its
accuracy improvements are
especially useful as the amount of training data decreases. In
Figures~\ref{fig:ow_two_tpr},~\ref{fig:ow_multi_tpr}, and~\ref{fig:ow_fpr},
we observe that
the gap between \attack and DF generally tends to increase as
the amount of data decreases. For instance, the Multi-TPR gap is 2.22\% with
1000 traces, 7.13\%
with 80 traces, and 13\% with 60 traces. At these same scales, the
FPR gap is 1.11\%, 2.21\%, and 3.12\%, respectively, and the Two-TPR gap is
2.13\%, 6.62\%, and
11.34\%, respectively. This indicates that an attacker with fewer capabilities
for collecting large amounts of training data has a better probability of both
fingerprinting a user while not falsely identifying them with \attack.
\S\ref{sec:why} relates this phenomenon to those observed in the
closed-world setting and provides a possible unifying explanation.

\subsubsection{Comparison against other attacks in the presence of WF defenses}
\label{sec:attack_comp}

\begin{table*}
    \caption{Open-world Multi-TPR and FPR for Var-CNN evaluated against three
    other state-of-the-art attacks.
    Evaluations are done in both natural settings (None) and WF defense
    settings against a constant-flow WF defense (Tamaraw) and a WF defense based on
    adaptive padding (WTF-PAD).
    We use 100 monitored sites, 100 monitored traces, 90000 unmonitored training
    sites, and 10000 unmonitored testing sites for this evaluation and simulate
    the attacks and defenses with the original code from each author.
    Note that Var-CNN gets higher Multi-TPR and lower FPR than any other attack,
    even in settings with WF defenses.
    Furthermore, note that Multi-TPR and FPR for Tamaraw is zero across
    nearly all attacks since the strong defense causes all attacks to classify
    every trace as unmonitored.
    The only exception is with $k$-FP, which we further explain
    in \S\ref{sec:attack_comp}.}
    \label{tab:comp_defense}
    \centering
    \begin{tabular}{c|cc|cccccccc}
        Defenses & \multicolumn{2}{c|}{Overhead} & \multicolumn{2}{c}{Var-CNN} &
        \multicolumn{2}{c}{DF~\cite{sirinam18}} & \multicolumn{2}{c}{$k$-FP~\cite{hayes16}}
        & \multicolumn{2}{c}{CUMUL~\cite{panchenko16}}\\
        & Bandwidth & Latency & Multi-TPR & FPR & Multi-TPR & FPR & Multi-TPR & FPR & Multi-TPR & FPR\\
        \midrule
        None & 0\% & 0\% & 89.2\% & 1.1\% & 88.4\% & 8.6\% & 70.4\% & 9.1\% & 85.6\% & 10.2\%\\
        Tamaraw~\cite{cai14_tam} & 63\% & 51\% & 0.0\% & 0.0\% & 0.0\% & 0.0\% & 4.1\% & 54.3\% & 0.0\% & 0.0\%\\
        WTF-PAD~\cite{juarez16} & 27\% & 0\% & 88.8\% & 0.7\% & 86.2\% & 5.4\% & 71.1\% & 9.5\% & 78.1\% & 12.6\%\\
    \end{tabular}
\end{table*}

While DF outperformed prior art on Sirinam et al.'s dataset~\cite{sirinam18},
these attacks have never been simultaneously tested on the Rimmer dataset~\cite{rimmer17}.
In this experiment, we evaluate Var-CNN and DF against prior art in
settings with and without WF defenses.
Note that here we exclude $k$-NN~\cite{wang14}
from our evaluation since it has been shown by several researchers to be
less accurate than both $k$-FP and CUMUL~\cite{hayes16, panchenko16, rimmer17, sirinam18}.

First, we assess our original assumption that DF is the current state-of-the-art
WF attack.
The first row in Table~\ref{tab:comp_defense} shows
Multi-TPR and FPR for all attacks in a medium-sized open-world with no WF defense.
Since DF has a higher Multi-TPR and lower FPR than all prior art, it
is still the current state-of-the-art attack, even on the Rimmer dataset~\cite{rimmer17}.
As noted by Sirinam et al.~\cite{sirinam18}, it also performs comparably
or slightly better than all prior art in WF defenses settings against
Tamaraw and WTF-PAD\@.

Having shown DF to be the best prior-art attack in settings with and without
WF defenses, we now assess Var-CNN's performance relative to DF\@.
In the undefended scenario, Var-CNN (with a confidence threshold of 0.5)
attains similar Multi-TPR to DF (with a confidence threshold of 0.7), 89.2\% compared
to 88.4\%.
However, it has a nearly $8\times$ lower FPR, going from 8.6\% to 1.1\%.
Thus, compared to Var-CNN, DF would incorrectly classify several more unmonitored websites
as being monitored, thereby falsely identifying users.

In the WF defense setting against Tamaraw and WTF-PAD, Var-CNN still retains
significant improvements over DF and the rest of the prior-art.
For example, with a Multi-TPR comparable to that of DF (88.8\% compared to 86.2\%, respectively),
Var-CNN achieves a nearly $8\times$ reduction in FPR from 5.4\% to 0.7\%.
Against other attacks such as $k$-FP and CUMUL, Var-CNN has even greater
improvements in both Multi-TPR and FPR\@.

Finally, we note that after trying several configurations of Tamaraw, we
were unable to get a configuration that yielded non-zero Multi-TPR and FPR
for all attacks.
While $k$-FP did manage to get a 4\% Multi-TPR against Tamaraw, the resulting
FPR of 54.3\% is too high to make reliable predictions.
Indeed, it appears that our Tamaraw configurations were so strong
that they caused most attacks to classify every instance as being unmonitored,
resulting in 0\% TPR and 0\% FPR\@.
In contrast, WTF-PAD (at least under the configuration we tested) sacrifices too
much security for low overheads.
As noted by Sirinam et al.~\cite{sirinam18}, it fails to protect against
attacks like Var-CNN and DF, which achieve relatively high TPR and relatively low FPR\@.

\section{Discussion}
\label{sec:discussion}
In this section, we discuss pertinent aspects of \attack.

\subsection{A note on runtimes}
Note that with some experimental settings, both DF and \attack
could possibly achieve better accuracies if allowed to train for longer.
However, this adds yet another parameter the attacker must optimize for
and introduces additional training time overheads.
We therefore made a simplifying assumption here to use
the same stopping scheme for each model across all experiments (see
Appendix~\ref{sec:model_design} for more details).

One of the weaknesses of \attack is that it takes a longer time
on average to train than DF due to the more complex
underlying model (ResNet-18), additional fully-connected layer for
basic cumulative features, and time model.
For example, for a closed-world with 100 monitored sites and 100 traces per
site, it took approximately 25 minutes to train Var-CNN using one GPU while it
took around 4 minutes to train DF. These runtimes scale roughly
proportionally with the amount of training data used.

While Var-CNN does cause increased runtimes, we believe they are
offset by its much improved performance in all settings tested.
For example, for DF to reach a comparable accuracy to Var-CNN in a closed-world
with 100 sites and 100 traces, it would need approximately $5\times$ the number
of traces (\S\ref{sec:closed_world}).
Moreover, we can parallelize the training procedure to decrease runtimes.
Since \attack direction and \attack time are independent models ensembled after
training, they can be trained separately on two GPUs.
In addition, recent work in Deep Learning research has shown that
models can be massively parallelized by using larger batch sizes, special
learning rate schemes, and more GPUs. A prime example of this is Goyal et
al.~\cite{goyal17}, who trained a ResNet-50 on a very large image
classification dataset in just one hour.
Previously, it would have taken several days to train on a single GPU.

\subsection{Why does \attack work in low-data settings?}
\label{sec:why}
As noted in \S\ref{sec:closed_world} and \S\ref{sec:ow_eval}, Var-CNN's accuracy
improvements over DF increase in the following scenarios:

\begin{enumerate}
\item
A fixed number of monitored sites and a decreasing number of monitored traces
in the closed-world.
\item
A fixed number of monitored traces and an increasing number of monitored sites
in the closed-world.
\item
A decreasing number of monitored traces and unmonitored training sites in the
open-world.
\end{enumerate}

In this section, we provide a possible general explanation for these trends and relate
them to the model techniques used in \attack.

For the following discussion, consider neural network optimization as finding a
good hypothesis (i.e., one that achieves
low training and generalization errors) among the overall
hypothesis space. All of the above settings make the
hypothesis space more vast and ambiguous to traverse.
For instance, decreasing
the number of monitored traces and unmonitored train sites (i.e.,
reducing the training data available to the classifier) provides less
information for the optimizer to find a good hypothesis with low generalization
error. On the other hand,
increasing the number of monitored sites as in the second scenario expands the overall
size of the hypothesis space with no additional optimization help from more
monitored traces. As the hypothesis space changes, a modern
convolutional neural network with enough weight parameters would still
likely achieve good training accuracy, as shown by Zhang et al.~\cite{zhang17}.
However, as the optimization landscape is more vast and ambiguous to traverse,
it would have an increasingly difficult time finding hypotheses that don't
overfit to the training data (i.e., that have low generalization error).

In contrast, one of the key benefits of ensembles, as noted by seminal
work~\cite{dietterich1990}, is statistical. When the hypothesis
space becomes unclear as with the above scenarios, ensembles help filter
out bad hypotheses that overfit to the training data, making the overall model
have better generalization capabilities. Given that \attack is composed of two
types of ensembles, one with an in-training combination of a dilated causal
ResNet-18 and cumulative features and the other with a post-training
ensemble of direction and timing models, we suspect that these ensembles
enable Var-CNN to perform especially well when the optimization landscape gets
more complex to navigate. Empirically, this was also observed in
\S\ref{sec:model_comp}, where the gap between the ensemble model and the best
performing constituent increases as the number of monitored traces decreases.

\section{Future work}
\label{sec:future_work}
Although \attack outperforms prior WF attacks in every setting tested,
especially with small amounts of training data, there are still many directions
where it could improve. In this section, we discuss limitations of
our work and possible avenues for future work.

\textbf{More powerful baseline models.}
Since deep learning is a rapidly accelerating field~\cite{lecun15},
applications of new model architecture breakthroughs could lead to
better results. For instance, here, we used the ResNet architecture as our
baseline CNN since it is
currently the most widely used state-of-the-art Image Classification CNN.
There are other Image Classification models such as larger variants of ResNets
or DenseNets~\cite{huang17} that could give better results. In our preliminary tests with
these architectures, however, we did not see significant-enough accuracy
improvements to justify their increased computational costs.

In addition, recent work on Synthetic Gradients~\cite{jaderberg17}
could lead to RNNs with the ability to train on much longer inputs
(e.g., the packet sequences used in this work).
Since RNNs were specifically made for temporal sequences,
this model might understand long-term packet interactions 
better than dilated causal convolutions.

Regardless of the architecture choice, we would like to note that nearly all of
the packet sequence classification insights used in this paper are
architecture-independent and can thus be applied to most future deep learning
attacks. For instance, dilated causal convolutions work with
any CNN architecture and ensembles with cumulative features and timing data
work with nearly all neural network models.

\textbf{Data augmentation.}
A common technique in Computer Vision research is to artificially expand the
training data size by using data augmentation---cropping, rotating, flipping,
shifting, and rescaling the image. This technique works in Computer Vision
because the artificial data is often similar-enough to
real-world data that it is useful to the model. Similar techniques for packet
sequences, such as shifting them a random number of packets one way or the
other, could be used to achieve even better low-data performance.

\textbf{User-sourced datasets.}
As described in \S\ref{sec:assumptions}, there are two main types of
assumptions, replicability and applicability. Both of these assumptions could
be tested in a real-world user trial study. Here, an adversary would draft a
number of real-world Tor users and monitor their packet sequences, sites
visited (including background traffic), and metadata settings (Tor versions,
circuit latencies, etc.). This would enable them to know how strong the WF
assumptions are in the real-world.

\textbf{Adversarial machine learning.}
Regular WF defenses such as recent work by Lu et al.~\cite{lu18} and
Wang et al.~\cite{wang17} degrade accuracy by blocking information leakage from
sources such as inter-packet timing, packet sequence length, and burst patterns.
However, more precise WF defenses could exist within the context of adversarial
attacks on machine learning models~\cite{goodfellow15}.
Assuming the WF defender has some knowledge of the WF attacker model, she might
be able to craft specialized perturbations to reduce the model's classification
ability while introducing less overhead than more traditional WF defenses.

One challenging problem here is defining a useful constraint for
adversarial attacks. Most work in adversarial machine learning has
focused on image classification, where $\ell_{p}$ norms constrain the total
amount of pixel-level perturbation allowed for an image.
With packet sequences, however, it is much harder to change individual inputs
as they must obey temporal orderings. For example, an outgoing packet cannot be
changed to incoming if that information is necessary in the future. The same
argument applies to changing the timestamp of a packet. Thus, creating a specialized
constraint for the set of all allowable perturbations will most likely be the
focus of future work.

Finally, as with all attack-defense paradigms, while adversarial machine
learning WF defenses might be able to initially defeat WF attacks, new
WF attacks could be trained to become robust against these defenses.
For instance, recent work by M\k{a}dry et al.~has shown that by
viewing adversarial machine learning within the context of convex optimization,
one can train a model that is robust to adversarial input~\cite{madry18}.
This would allow for a WF attack to be resistant to adversarial WF defense perturbations.

\textbf{Code release.}
To support future work, we have made our code, including our re-implementation of
DF, publicly available at \url{https://github.com/sanjit-bhat/Var-CNN}.
Please refer to Rimmer et al.'s paper~\cite{rimmer17} for instructions
on downloading their dataset.

\section{Conclusion}
\label{sec:conclusion}
In this work, we present \attack, a novel website fingerprinting
attack that combines a strong baseline ResNet CNN model with several powerful
insights for packet sequence classification: dilated causal convolutions,
automatically extracted direction and timing features, and manually extracted
cumulative features. In open-world settings with large amounts of data,
\attack achieves over 1\% better TPR and $4\times$ lower FPR than prior-art.
In addition, Var-CNN's
improvements are especially relevant in low-data scenarios, where deep learning
models typically suffer. Here, it reduces prior-art FPR by 3.12\% while
increasing TPR by 13\%.

Overall, Var-CNN's model insights, which can be applied to most future neural
network models, allow it to need less training data than prior art.
This lowers the likelihood of data staleness performance issues and allows
a weaker attacker with fewer data collection resources to successfully
perform a powerful WF attack.

\section{Acknowledgements}
This work was done as part of the MIT PRIMES program while Sanjit Bhat and
David Lu were students at Acton-Boxborough Regional High School.
The authors were partially supported by National Science
Foundation grant No.~1813087.
The authors would like to thank Dimitris Tsipras and the M\k{a}dry Lab at
MIT for providing some of the compute resources used to run these experiments.

\bibliographystyle{plain}
{\footnotesize \bibliography{references}}

\begin{thebibliography}{10}

\bibitem{alexa_topsites}
{The Top 500 Sites on the Web}.
\newblock \url{https://www.alexa.com/topsites}, 2017.

\bibitem{tensorflow15}
Mart{\'{\i}}n Abadi, Ashish Agarwal, Paul Barham, Eugene Brevdo, Zhifeng Chen,
  Craig Citro, Gregory~S. Corrado, Andy Davis, Jeffrey Dean, Matthieu Devin,
  Sanjay Ghemawat, Ian~J. Goodfellow, Andrew Harp, Geoffrey Irving, Michael
  Isard, Yangqing Jia, Rafal J{\'{o}}zefowicz, Lukasz Kaiser, Manjunath Kudlur,
  Josh Levenberg, Dan Man{\'{e}}, Rajat Monga, Sherry Moore, Derek~Gordon
  Murray, Chris Olah, Mike Schuster, Jonathon Shlens, Benoit Steiner, Ilya
  Sutskever, Kunal Talwar, Paul~A. Tucker, Vincent Vanhoucke, Vijay Vasudevan,
  Fernanda~B. Vi{\'{e}}gas, Oriol Vinyals, Pete Warden, Martin Wattenberg,
  Martin Wicke, Yuan Yu, and Xiaoqiang Zheng.
\newblock {TensorFlow: Large-Scale Machine Learning on Heterogeneous Systems}.
\newblock {\em arXiv preprint arXiv:1603.04467}, 2015.

\bibitem{abe16}
Kota Abe and Shigeki Goto.
\newblock {Fingerprinting Attack on Tor Anonymity using Deep Learning}.
\newblock In {\em Proceedings of the Asia-Pacific Advanced Network Research
  Workshop}, volume~42, pages 15--20, 2016.

\bibitem{bissias06}
George~D. Bissias, Marc Liberatore, David Jensen, and Brian~N. Levine.
\newblock {Privacy Vulnerabilities in Encrypted HTTP Streams}.
\newblock {\em Privacy Enhancing Technologies}, pages 1--11, 2006.

\bibitem{cai14_tam}
Xiang Cai, Rishab Nithyanand, Tao Wang, Rob Johnson, and Ian Goldberg.
\newblock {A Systematic Approach to Developing and Evaluating Website
  Fingerprinting Defenses}.
\newblock In {\em Proceedings of the ACM Conference on Computer and
  Communications Security}, pages 227--238, 2014.

\bibitem{cai12}
Xiang Cai, Xin~C. Zhang, Brijesh Joshi, and Rob Johnson.
\newblock {Touching from a Distance: Website Fingerprinting Attacks and
  Defenses}.
\newblock In {\em Proceedings of the ACM Conference on Computer and
  Communications Security}, pages 605--616, 2012.

\bibitem{cheng98}
Heyning Cheng and Ron Avnur.
\newblock {Traffic Analysis of SSL Encrypted Web Browsing}.
\newblock
  \url{https://pdfs.semanticscholar.org/1a98/7c4fe65fa347a863dece665955ee7e01791b.pdf},
  1998.

\bibitem{chollet2015}
Fran\c{c}ois Chollet et~al.
\newblock {Keras}.
\newblock \url{https://keras.io}, 2015.

\bibitem{tor_metric}
Tor Developers.
\newblock Tor metrics portal.
\newblock https://metrics.torproject.org, 2018.

\bibitem{dietterich1990}
Thomas~G. Dietterich.
\newblock {Ensemble Methods in Machine Learning}.
\newblock In {\em Proceedings of the International Workshop on Multiple
  Classifier Systems}, 2000.

\bibitem{tor}
Roger Dingledine, Nick Mathewson, and Paul Syverson.
\newblock {Tor: The Second-Generation Onion Router}.
\newblock In {\em Proceedings of the 13th USENIX Security Symposium}, pages
  303--320, 2004.

\bibitem{dyer12}
Kevin~P. Dyer, Scott~E. Coull, Thomas Ristenpart, and Thomas Shrimpton.
\newblock {Peek-a-Boo, I Still See You: Why Efficient Traffic Analysis
  Countermeasures Fail}.
\newblock In {\em Proceedings of the IEEE Symposium on Security and Privacy},
  pages 332--346, 2012.

\bibitem{goodfellow15}
Ian~J. Goodfellow, Jonathon Shlens, and Christian Szegedy.
\newblock {Explaining and Harnessing Adversarial Examples}.
\newblock In {\em Proceedings of the International Conference on Learning
  Representations}, 2015.

\bibitem{goyal17}
Priya Goyal, Piotr Doll\'{a}r, Ross Girshick, Pieter Noordhuis, Lukasz
  Wesolowski, Aapo Kyrola, Andrew Tulloch, Yangqing Jia, and Kaiming He.
\newblock {Accurate, Large Minibatch SGD: Training ImageNet in 1 Hour}.
\newblock {\em arXiv preprint arXiv:1706.02677}, 2017.

\bibitem{gupta17}
Ankit Gupta and Alexander~M. Rush.
\newblock {Dilated Convolutions for Modeling Long-Distance Genomic
  Dependencies}.
\newblock In {\em Proceedings of the 34th International Conference on Machine
  Learning, Workshop on Computational Biology}, 2017.

\bibitem{hayes16}
Jamie Hayes and George Danezis.
\newblock {$k$-fingerprinting: A Robust Scalable Website Fingerprinting
  Technique}.
\newblock In {\em Proceedings of the 25th USENIX Security Symposium}, pages
  1187--1203, 2016.

\bibitem{resnet}
Kaiming He, Xiangyu Zhang, Shaoqing Ren, and Jian Sun.
\newblock {Deep Residual Learning for Image Recognition}.
\newblock {\em arXiv preprint arXiv:1512.03385}, 2015.

\bibitem{herrmann09}
Dominik Herrmann, Rolf Wendolsky, and Hannes Federrath.
\newblock {Website Fingerprinting: Attacking Popular Privacy Enhancing
  Technologies with the Multinomial Naïve-Bayes Classifier}.
\newblock In {\em Proceedings of the ACM Workshop on Cloud Computing Security},
  pages 31--42, 2009.

\bibitem{hintz03}
Andrew Hintz.
\newblock {Fingerprinting Websites Using Traffic Analysis}.
\newblock {\em Privacy Enhancing Technologies}, pages 171--178, 2003.

\bibitem{hochreiter1997}
Sepp Hochreiter and J\"{u}rgen Schmidhuber.
\newblock {Long Short-Term Memory}.
\newblock {\em Neural Computation}, 9(8):1735--1780, 1997.

\bibitem{huang17}
Gao Huang, Zhuang Liu, Laurens van~der Maaten, and Kilian~Q. Weinberger.
\newblock Densely connected convolutional networks.
\newblock In {\em Proceedings of the IEEE Conference on Computer Vision and
  Pattern Recognition}, 2017.

\bibitem{ioffe15}
Sergey Ioffe and Christian Szegedy.
\newblock {Batch Normalization: Accelerating Deep Network Training by Reducing
  Internal Covariate Shift}.
\newblock In {\em Proceedings of the 32nd International Conference on Machine
  Learning}, 2015.

\bibitem{jaderberg17}
Max Jaderberg, Wojciech~M. Czarnecki, Simon Osindero, Oriol Vinyals, Alex
  Graves, David Silver, and Koray Kavukcuoglu.
\newblock {Decoupled Neural Interfaces using Synthetic Gradients}.
\newblock In {\em Proceedings of the 34th International Conference on Machine
  Learning}, 2017.

\bibitem{juarez14}
Marc Juarez, Sadia Afroz, Gunes Acar, Claudia Diaz, and Rachel Greenstadt.
\newblock {A Critical Evaluation of Website Fingerprinting Attacks}.
\newblock In {\em Proceedings of the ACM Conference on Computer and
  Communications Security}, 2014.

\bibitem{juarez16}
Marc Juarez, Mohsen Imani, Mike Perry, Claudia Diaz, and Matthew Wright.
\newblock {Toward an Efficient Website Fingerprinting Defense}.
\newblock In {\em Proceedings of the European Symposium on Research in Computer
  Security}, pages 27--46, 2016.

\bibitem{kingma2014}
Diederik~P. Kingma and Jimmy Ba.
\newblock {Adam: A Method for Stochastic Optimization}.
\newblock In {\em Proceedings of the 3rd International Conference on Learning
  Representations}, 2015.

\bibitem{krizhevsky2012}
Alex Krizhevsky, Ilya Sutskever, and Geoffrey~E. Hinton.
\newblock {ImageNet Classification with Deep Convolutional Neural Networks}.
\newblock In {\em Proceedings of the Conference on Neural Information
  Processing Systems}, pages 1097--1105, 2012.

\bibitem{lecun15}
Yann LeCun, Yoshua Bengio, and Geoffrey~E. Hinton.
\newblock {Deep Learning}.
\newblock {\em Nature}, 521:436--444, 2015.

\bibitem{lecun1998}
Yann LeCun, Leon Bottou, Yoshua Bengio, and Patrick Haffner.
\newblock {Gradient-Based Learning Applied to Document Recognition}.
\newblock {\em Proceedings of the IEEE}, 86(11):2278--2324, 1998.

\bibitem{liberatore06}
Marc Liberatore and Brian~N. Levine.
\newblock {Inferring the Source of Encrypted HTTP Connections}.
\newblock In {\em Proceedings of the 13th ACM Conference on Computer and
  Communications Security}, pages 255--263, 2006.

\bibitem{lu18}
David Lu, Sanjit Bhat, Albert Kwon, and Srinivas Devadas.
\newblock {DynaFlow: An Efficient Website Fingerprinting Defense Based on
  Dynamically-Adjusting Flows}.
\newblock In {\em Proceedings of the ACM Workshop on Privacy in the Electronic
  Society}, 2018.

\bibitem{lu10}
Liming Lu, Ee-Chien Chang, and Mun~C. Chan.
\newblock {Website Fingerprinting and Identification Using Ordered Feature
  Sequences}.
\newblock In {\em Proceedings of the European Symposium on Research in Computer
  Security}, pages 199--214, 2010.

\bibitem{madry18}
Aleksander M\k{a}dry, Aleksandar Makelov, Ludwig Schmidt, Dimitris Tsipras, and
  Adrian Vladu.
\newblock {Towards Deep Learning Models Resistant to Adversarial Attacks}.
\newblock In {\em Proceedings of the International Conference on Learning
  Representations}, 2018.

\bibitem{panchenko16}
Andriy Panchenko, Fabian Lanze, Aandreas Zinnen, and Martin Henze.
\newblock {Website Fingerprinting at Internet Scale}.
\newblock In {\em Proceedings of the 16th Network and Distributed System
  Security Symposium}, 2016.

\bibitem{panchenko11}
Andriy Panchenko, Lukas Niessen, Andreas Zinnen, and Thomas Engel.
\newblock {Website Fingerprinting in Onion Routing Based Anonymization
  Networks}.
\newblock In {\em Proceedings of the ACM Workshop on Privacy in the Electronic
  Society}, pages 103--114, 2011.

\bibitem{rimmer17}
Vera Rimmer, Davy Preuveneers, Marc Juarez, Tom~V. Goethem, and Wouter Joosen.
\newblock {Automated Feature Extraction for Website Fingerprinting through Deep
  Learning}.
\newblock In {\em Proceedings of the Network and Distributed System Security
  Symposium}, 2018.

\bibitem{simonyan14}
Karen Simonyan and Andrew Zisserman.
\newblock {Very Deep Convolutional Networks for Large-Scale Image Recognition}.
\newblock {\em arXiv preprint arXiv:1409.1556}, 2014.

\bibitem{sirinam18}
Payap Sirinam, Mohsen Imani, Marc Juarez, and Matthew Wright.
\newblock {Deep Fingerprinting: Undermining Website Fingerprinting Defenses
  with Deep Learning}.
\newblock In {\em Proceedings of the ACM Conference on Computer and
  Communications Security}, 2018.

\bibitem{srivastava14}
Nitish Srivastava, Geoffrey~H. Hinton, Alex Krizhevsky, Ilya Sutskever, and
  Ruslan Salakhutdinov.
\newblock {Dropout: A Simple Way to Prevent Neural Networks from Overfitting}.
\newblock {\em Journal of Machine Learning Research}, 15:1929--1958, 2014.

\bibitem{sun02}
Qixiang Sun, Daniel~R. Simon, Yi-Min Wang, Wilf Russell, Venkata~N.
  Padmanabhan, and Lili Qiu.
\newblock {Statistical Identification of Encrypted Web Browsing Traffic}.
\newblock In {\em Proceedings of the IEEE Symposium on Security and Privacy},
  pages 19--30, 2002.

\bibitem{inception}
Christian Szegedy, Sergey Ioffe, Vincent Vanhoucke, and Alex Alemi.
\newblock {Inception-v4, Inception-ResNet and the Impact of Residual
  Connections on Learning}.
\newblock {\em arXiv preprint arXiv:1602.07261}, 2016.

\bibitem{oord16}
Aaron van~den Oord, Sander Dieleman, Heiga Zen, Karen Simonyan, Oriol Vinyals,
  Alex Graves, Nal Kalchbrenner, Andrew Senior, and Koray Kavukcuoglu.
\newblock {WaveNet: A Generative Model for Raw Audio}.
\newblock {\em arXiv preprint arXiv:1609.03499}, 2016.

\bibitem{wang14}
Tao Wang, Xiang Cai, Rob Johnson, and Ian Goldberg.
\newblock {Effective Attacks and Provable Defenses for Website Fingerprinting}.
\newblock In {\em Proceedings of the 23rd USENIX Security Symposium}, pages
  143--157, 2014.

\bibitem{wang13}
Tao Wang and Ian Goldberg.
\newblock {Improved Website Fingerprinting on Tor}.
\newblock In {\em Proceedings of the ACM Workshop on Privacy in the Electronic
  Society}, 2013.

\bibitem{wang16}
Tao Wang and Ian Goldberg.
\newblock {On Realistically Attacking Tor with Website Fingerprinting}.
\newblock In {\em Proceedings on Privacy Enhancing Technologies}, pages 21--36,
  2016.

\bibitem{wang17}
Tao Wang and Ian Goldberg.
\newblock {Walkie-Talkie: An Efficient Defense Against Passive Website
  Fingerprinting Attacks}.
\newblock In {\em Proceedings of the USENIX Security Symposium}, pages
  1375--1390, 2017.

\bibitem{yu16}
Fisher Yu and Vladlen Koltun.
\newblock {Multi-Scale Context Aggregation By Dilated Convolutions}.
\newblock In {\em Proceedings of the International Conference on Learning
  Representations}, 2016.

\bibitem{zhang17}
Chiyuan Zhang, Samy Bengio, Moritz Hardt, Benjamin Recht, and Oriol Vinyals.
\newblock {Understanding Deep Learning Requires Rethinking Generalization}.
\newblock In {\em Proceedings of the International Conference on Learning
  Representations}, 2017.

\end{thebibliography}

\appendix
\appendixpage
\section{\attack model design}
\label{sec:model_design}
To come up with the final \attack model, we performed several different
experiments testing various aspects of the model design. Here, we mention a few
important results. For full parameter choices, please refer to our code.

\subsection{Ensemble schemes}
One of our main problems was deciding how to most effectively combine the
direction, time, and metadata models. First, note that each model yielded
varying standalone performance. For instance, in a closed-world with 100 sites
and 90 monitored traces, both the direction and time models achieved above
90\% accuracy while the metadata model consistently got 35\% accuracy.
If we just averaged all three models post-training, this caused an accuracy
\emph{decrease} since the metadata model was significantly worse than the other
two. This finding led us to use metadata during training rather than after. In
our tests, we noticed that adding metadata to both the direction and time
ResNets acted as a stabilizer, reducing stochasticity and improving final
accuracy. Intuitively, the overall model learned when to use the
automatically extracted ResNet features over the manually extracted
cumulative features, and vice-versa, balancing each source of information.

Now that we knew to combine the ResNet with the cumulative features in-training,
one remaining question was how to combine the direction and timing information.
In early experiments, we tried to use in-training weight-sharing techniques.
While this certainly did not double the number of model parameters, it resulted
in an accuracy drop compared to the individual direction and timing models.
This was likely due to the dissimilarity between direction
and timing inputs (e.g., on a meta-level, the former is sampled from a discrete
distribution while the latter from a continuous distribution).
As a result, features learned from direction inputs were markedly different
from those learned from timing inputs, and the shared model struggled to find
a set of shared weights.

Unfortunately, the alternative to weight sharing---an in-training combination
of two separate direction and timing models---often led to
reduced accuracy as the fundamental optimization
problem got much harder with twice the number of parameters.
Instead, we ensembled these models post-training by combining their
softmax outputs. In our experiments, we compared various schemes for combining,
such as using the validation set to find a weighted average and doing simple
averaging, against a simple sweep to find the optimal weight. In most cases,
our results indicated that the first method was overfitting the validation set,
whereas a simple average usually achieved near-optimal accuracy. Therefore,
for the final \attack model we performed a simple average over the softmax
outputs of the direction and time models trained jointly with metadata, as
shown in Figure~\ref{fig:ensemble}.

\subsection{Learning rate decay and early stopping}
\label{sec:val_params}

Rather than having a fixed learning rate drop-off scheme with a fixed
number of epochs, we found it much more effective to decide these based
on validation set performance. Specifically, we started training with a
learning rate of 0.001, the default value for the Adam
optimizer~\cite{kingma2014}, and allowed the network to train for 5 epochs
without improving validation accuracy before we reduced learning rate by a
factor of $\sqrt{0.1}$ (i.e., $new\textunderscore lr = \sqrt{0.1} *
old\textunderscore lr$). The minimum possible
learning rate was 0.00001, and we stopped training after not improving
validation accuracy for 10 epochs, twice the number used for learning rate
decay. We saved the best performing model on the validation set and
reloaded this model to perform final classification on the test set.

Two parameters we experimented with here were the initial learning rate and the
base patience used for learning rate decay. From our experiments, reducing
learning rate from the default resulted in higher local minima in the loss
landscape, while increasing learning rate did not improve model accuracy.
Increasing base patience sometimes allowed for slightly better accuracy, but it
also significantly increased the average number of epochs for training. With a
base patience of 5, both the direction and time models used between 30--50
epochs across all experimental settings.

\subsection{Dropout and $\ell_{2}$ decay}
With both the direction and time models, training accuracy reached around
99.99\% by the end of training, indicating that the model may have overfit to the
training data. To curb this, we tried various mechanisms, such as introducing
Dropout~\cite{srivastava14} and $\ell_{2}$ decay to the ResNet model and
the final fully-connected layer after concatenating the ResNet and metadata
models. We observed that $\ell_{2}$ decay did not manage to reduce
overfitting on the ResNet. However, using a Dropout rate of 0.5 on the final
fully-connected layer did reduce overfitting to some extent, decreasing the
generalization error.

\end{document}